\documentclass[prd,%
preprint,%
tightenlines,
showpacs,nofootinbib,eqsecnum,amsfonts,amsmath,amssymb]{revtex4}
\usepackage{bm}
\usepackage{graphicx}

\allowdisplaybreaks

\begin{document}

\title{Higher-order spin effects in the dynamics of compact binaries
\\II. Radiation field}

\author{Luc Blanchet$^{a}$, Alessandra Buonanno$^{b,c,a}$ and Guillaume
Faye$^{a}$} \affiliation{$^a$ ${\mathcal{G}}{\mathbb{R}}
\varepsilon{\mathbb{C}}{\mathcal{O}}$, Institut d'Astrophysique de
Paris, UMR 7095 CNRS Universit\'e Pierre \& Marie Curie, 98$^{\text{bis}}$
boulevard Arago, 75014 Paris,
France\\ $^{b}$ Department of Physics, University of Maryland, College
Park, MD 20742 \\ $^c$ AstroParticule et Cosmologie (APC), UMR
7164-CNRS, 11, place Marcellin Berthelot, 75005 Paris, France}

\begin{abstract}
Motivated by the search for gravitational waves emitted by binary black
holes, we investigate the gravitational radiation field of point
particles with spins within the framework of the
multipolar-post-Newtonian wave generation formalism. We compute: (i) the
spin-orbit (SO) coupling effects in the binary's mass and current
quadrupole moments one post-Newtonian (1PN) order beyond the dominant
effect, (ii) the SO contributions in the gravitational-wave energy flux
and (iii) the secular evolution of the binary's orbital phase up to
2.5PN order. Crucial ingredients for obtaining the 2.5PN contribution in
the orbital phase are the binary's energy and the spin precession
equations, derived in paper~I of this series. These results provide more
accurate gravitational-wave templates to be used in the data analysis of
rapidly rotating Kerr-type black-hole binaries with the ground-based
detectors LIGO, Virgo, GEO 600 and TAMA300, and the space-based detector LISA.
\end{abstract}


\maketitle

\section{Introduction}\label{secI}

The aim of this paper is to derive the spin-orbit coupling terms in the
gravitational radiation field of compact binary systems one
post-Newtonian (1PN)\,\footnote{By $n$PN we refer to the terms of
relative order $(v/c)^{2n}$ where $v$ is the binary's orbital velocity
and $c$ the speed of light.} order beyond the dominant effect. Paper I
of this series~\cite{FBB06spin} dealt with the problem of the spin-orbit
contributions in the compact binary equations of motion at 1PN relative
order.

Our motivation is the on going search for gravitational waves (GWs) emitted by
inspiralling binary systems of \textit{spinning}, and possibly maximally
spinning, black holes in the network of detectors LIGO (Laser Interferometer
Gravitational Wave Observatory), Virgo, GEO 600 and TAMA300, and the future
search with the space-based detector LISA. When the Kerr black holes are
maximally spinning (or close to maximal), the GW templates need to take into
account the effects of spins, not only for an accurate parameter
estimation~\cite{CF94,PW95,KKS95,C98,V04,BBW05}, but also for a successful
detection~\cite{3mn,ACST94,A95,A96a,A96b,GKV03,BCV03b,PBCV04,
  GK03,BCPV04,GIKB04,BCPTV05}. Furthermore, spin effects should be included at
PN orders beyond the currently known dominant spin-orbit and spin-spin terms.
The contributions of spins are added to the templates developed for the case
of non-spinning binary black holes or neutron stars, and which are currently
known at 3.5PN order~\cite{BFIJ02,Bliving, ABIQ04,BDEI04}. The spins represent
some of the possible effects depending on the internal structure of the bodies
which can be numerically important in the LIGO/Virgo bandwidth. This is true
even if we observe only the inspiral phase of moderate-mass black holes with
individual mass less than 10 $M_\odot$. Within the PN formalism the compact
objects are treated as point particles. It is then
natural to model spinning black holes as point particles with spins.

The equations of motion including the spin-orbit (SO) effect were
obtained in paper~I at 1PN relative order, which corresponds formally to
the 2.5PN order beyond the Newtonian force law in the case of maximally
rotating compact objects. Paper I essentially confirmed the equations of
motion derived previously by Tagoshi, Ohashi and Owen~\cite{TOO01}.
Furthermore, paper~I derived the complete set of Noetherian conserved
integrals of motion at that order (namely, 2.5PN for the spins). In the
present paper, we tackle the problem of the gravitational radiation
field at the same 2.5PN order, using the multipolar PN wave generation
formalism of Refs.~\cite{BD86,B87,BD89,B95,B96,B98mult}. More precisely,
we shall compute here the SO contributions in the compact binary's
\textit{mass-type} and \textit{current-type} quadrupole moments, both of
them with 1PN relative accuracy. These moments, together with some
easily computed higher multipole moments which necessitate only the
lowest-order precision, are necessary to compute the total GW energy
flux $\mathcal{F}$. The computation of the \textit{current} quadrupole
moment was previously attempted in Ref.~\cite{OTO98}, but we shall point
out two important flaws in that reference (see below for details). Our
result for the current quadrupole moment is substantially different from
that of~\cite{OTO98}. Concerning the \textit{mass} quadrupole moment, it
is computed here for the first time. Having in hand the total energy
flux, using the center-of-mass energy $E$ computed in paper~I, we deduce
(by energy balance arguments) the equation of secular evolution of the
binary's orbital frequency. The latter is the crucial ingredient needed
to build GW templates for spinning compact binaries.

To describe particles with spins we use the formalism originally developed in
Ref.~\cite{Papa51spin,BOC75,BOC79,BI80} (an effective field theory scheme has
recently been proposed \cite{GR06,Porto06,PR06}). This formalism has already
been successfully applied to the problem of spinning compact binaries in
Refs.~\cite{KWW93,K95,OTO98,cho98,TOO01,Ger99,MVGer05}, and, in the test-mass
limit case, in Refs.~\cite{MST96,TMSS96}. In particular, Kidder, Will and
Wiseman derived in Ref.~\cite{KWW93,K95} the lowest-order spin-coupling
effects --- at 1.5PN order in the case of maximal Kerr black holes ---, and
the first spin-spin effect, quadratic in the spins --- appearing at 2PN order
---, in the equations of motion and the gravitational radiation field. As we
shall see below we find complete agreement at that order with their results.
The present paper together with paper~I extend therefore the
works~\cite{KWW93,K95} to include the next-order spin effects. Since those
effects, of 2.5PN order, are linear in the spins (the next-order spin-spin
term coming along at 3PN order), we complete the derivation of all the spin
contributions in the GW form up to 2.5PN order.
 
The paper is organized as follows. In Sec.~\ref{secII} we review the
general formalism for wave generation from arbitrary PN sources. In
Sec.~\ref{secIII} we compute the multipole moments of compact binary
systems at the lowest PN level in the spins (which means 1.5PN for
mass moments, and 0.5PN for current ones). Sec.~\ref{secIV}
constitutes the core of the paper. We compute there the mass and
current quadrupole moments at 1PN relative order, \textit{i.e.} 2.5PN
and 1.5PN for the mass and current types, respectively. In particular,
the crucial contribution of non-compact-support terms (which are
sourced by the gravitational field itself) is obtained. The final
results for the multipole moments and the GW flux are presented in
Sec.~\ref{secV}, and in Sec.~\ref{secVI}, we reduce those results to
the physically relevant case of quasi-circular orbits. In
Sec.~\ref{secVII}, we express the main equations defining the binary
evolution and the GW signal in terms of spin variables with constant
magnitude, which generalize those used in Refs.~\cite{KWW93,K95}. In
Sec.~\ref{secVIII}, relying notably on the results of paper~I, we
obtain the secular evolution of the orbital frequency in the case of
circular orbits and discuss some implications for ground-based and
space-based detectors. Section~\ref{secIX} summarizes our main
conclusions.

All the notations and conventions are the same as in paper~I. Notably,
in order to explicitly display the powers of $1/c$ appropriate
for maximally spinning compact objects --- in which case the spin variable
is formally of the order 0.5PN $\sim 1/c$ ---, it is convenient to adopt
as basic spin variable a quantity having the dimension of an angular
momentum multiplied by $c$, namely $S=c\,S^\mathrm{true}$ (see paper~I for 
discussion). The precise definition of the spin contravariant vector we
use and the supplementary condition it satisfies are given in Sec.~II
of paper~I. See the same reference for the details concerning the
stress-energy tensor of spinning point particles.

\section{Post-Newtonian source multipole moments}\label{secII}

We start with a brief review of the PN multipole moment formalism at the basis
of this approach (full details can be found in
Refs.~\cite{B95,B96,B98mult,Bliving}). This formalism is valid for general
localized material sources, satisfying the usual PN requirements of weak
self-gravity, slow motion and weak internal stresses. In particular, the size
of the source $a$ has to be small with respect to the typical (reduced)
wavelength $\lambda\!\!\!\!\mathop{}^{-}$ of the gravitational radiation this
source produces, \emph{i.e.}
$a/\lambda\!\!\!\!\mathop{}^{-}=\mathcal{O}(\epsilon)$, with $\epsilon\sim
v/c$ being the slowness PN parameter. We shall abbreviate it as $\epsilon=1/c$
and denote the PN remainder terms by $\mathcal{O}(1/c^n)$ henceforth.

Let $x^\mu = (c\,t,\mathbf{x})$ be an harmonic coordinate system
covering the whole material source. We pose $h^{\mu\nu}\equiv
\sqrt{-g}\, g^{\mu\nu}-\eta^{\mu\nu}$ where $g^{\mu\nu}$ and $g$ are the
inverse and the determinant of the usual covariant metric $g_{\mu\nu}$,
and where $\eta^{\mu\nu}$ denotes an auxiliary Minkowskian background
metric (Greek letters represent space-time indices; our signature is
$+2$). The Einstein field equations, relaxed by the harmonic-coordinate
condition, $\partial_\nu h^{\mu\nu} = 0$, take the form
\begin{equation}\label{EE} \Box \,h^{\mu\nu} = \frac{16\pi
G}{c^4}\,\tau^{\mu\nu} \equiv \frac{16\pi G}{c^4}\,\vert g\vert
\,T^{\mu\nu}+\Lambda^{\mu\nu}\left[h\right]\,,
\end{equation}
where $\Box\equiv\eta^{\mu\nu}\partial_{\mu\nu}$ is the flat d'Alembertian,
and where the second equality defines the stress-energy \textit{pseudo tensor}
$\tau^{\mu\nu}$ of the matter and gravitational fields in harmonic
coordinates. Here $T^{\mu\nu}$ is the stress-energy tensor of the matter
fields and $\Lambda^{\mu\nu}\left[h\right]$ represents the gravitational
source term, namely a complicated non-linear functional of $h^{\rho\sigma}$
and its space-time derivatives $\partial_\lambda h^{\rho\sigma}$ (see
\textit{e.g.}~\cite{Bliving} for the expression). We shall see that
$\Lambda^{\mu\nu}$ gives a crucial contribution to 
the multipole moments at the relative 1PN order in the spins. The
stress-energy pseudo tensor is conserved by virtue of the harmonic-coordinate
condition,
\begin{equation}\label{conservelaw}
\partial_\nu h^{\mu\nu} = 0~~\Longrightarrow~~\partial_\nu\tau^{\mu\nu}
= 0\,.
\end{equation}

The multipole moments of the source are generated by the
components of the pseudo tensor $\tau^{\mu\nu}$ or, more precisely, of
its formal PN expansion
$\overline{\tau}^{\mu\nu}\equiv\mathrm{PN}[\tau^{\mu\nu}]$. In this
sense, the formalism is physically valid for PN sources only. The PN
expansion $\overline{\tau}^{\mu\nu}$ has a special structure which can
be matched to the exterior multipolar field of a PN source, allowing one
to define an appropriate notion of PN multipole
moments~\cite{B95,B96,B98mult}. It is convenient to define (Latin
letters representing space indices)
\begin{subequations}\label{Sigma}\begin{eqnarray}
&&\Sigma \equiv c^{-2}\left[\overline{\tau}^{00}
+\overline{\tau}^{ii}\right]~~\text{where
$\overline{\tau}^{ii}=\delta_{ij}\overline{\tau}^{ij}$}\,,\\ &&\Sigma_i
\equiv c^{-1}\,\overline{\tau}^{0i}\,,\\ &&\Sigma_{ij} \equiv
\overline{\tau}^{ij}\,.
\end{eqnarray}\end{subequations}
The mass-type moments $I_L(t)$ and current-type ones $J_L(t)$ are referred to
as the \textit{source} multipole moments in order to distinguish them from the
so-called \textit{radiative} moments, seen at infinity and generally denoted
$U_L(t)$ and $V_L(t)$. They are given by\,\footnote{In our notation, $L\equiv
  i_1\cdots i_\ell$ represents a multi-index composed of $\ell$ multipolar
  indices $i_1, \cdots, i_\ell$, and $x_L\equiv x_{i_1}\cdots x_{i_\ell}$
  stands for the product of $\ell$ spatial vectors $x^i\equiv x_i$. We denote
  the symmetric-trace-free (STF) projection by means of a hat over the tensor
  symbol, $\hat{x}_L\equiv\mathrm{STF}(x_{i_1}\cdots x_{i_\ell})$, or
  of brackets $\langle\rangle$ surrounding the indices, $\hat{x}_L\equiv
  x_{\langle L\rangle}$. When the multi-indices $L$ are summed up (dummy indices),
  we omit to write the $\ell$ summation symbols from 1 to 3 over their
  indices.}
\begin{subequations}\label{ILJL}\begin{eqnarray}
I_L(t)&=& \mathop{\mathrm{FP}}_{B=0}\,\int
d^3\mathbf{x}\,\vert\mathbf{x}\vert^B\int^1_{-1} dz\left\{
\delta_\ell(z)\,\hat{x}_L\,\Sigma
-\frac{4(2\ell+1)}{c^2(\ell+1)(2\ell+3)} \,\delta_{\ell+1}(z)
\,\hat{x}_{iL} \,\dot{\Sigma}_i\right.\nonumber\\ &&\qquad\quad \left.
+\frac{2(2\ell+1)}{c^4(\ell+1)(\ell+2)(2\ell+5)}
\,\delta_{\ell+2}(z)\,\hat{x}_{ijL}\ddot{\Sigma}_{ij}\right\}
(\mathbf{x},t+z \vert{\mathbf{x}}\vert/c)\,,\\J_L(t)&=&
\mathop{\mathrm{FP}}_{B=0}\,\varepsilon_{ab\langle i_\ell} \int d^3
\mathbf{x}\,\vert\mathbf{x}\vert^B \int^1_{-1} dz\biggl\{
\delta_\ell(z)\,\hat{x}_{L-1\rangle a} \,\Sigma_b \nonumber\\
&&\qquad\quad -\frac{2\ell+1}{c^2(\ell+2)(2\ell+3)}
\,\delta_{\ell+1}(z)\,\hat{x}_{L-1\rangle ac}
\,\dot{\Sigma}_{bc}\biggr\} (\mathbf{x},t+z \vert\mathbf{x}\vert/c)\,.
\end{eqnarray}\end{subequations}
These expressions are general, in the sense that they are formally valid at
any PN order, so that there are no remainder terms $\mathcal{O}(c^{-n})$
involved. The dots indicate the time derivatives and $\varepsilon_{abc}$
is the Levi-Civita symbol in 3 dimensions with $\varepsilon_{123}=1$. Notice the
peculiar feature that, besides the usual spatial integration, the moments
involve an extra integral over a variable $z$ defining a ``cone'' of
integration $u=t+z \vert\mathbf{x}\vert/c$; hence the
sources $\Sigma_{\mu\nu}$ depend on the point $(\mathbf{x},u)$ as
indicated. The ``weighting'' function associated with the $z$-integral,
$\delta_\ell (z) = \frac{(2\ell+1)!!}{2^{\ell+1} \ell!}
\,(1-z^2)^\ell$, is normalized in such a way that $\int_{-1}^1
dz\,\delta_\ell (z) =1$. When performing explicitly the PN
expansion of the moments, the $z$-integration is to be transformed into
an infinite \textit{local} series (in the sense that $z$ is integrated
out), which constitutes the basis of the practical evaluation of the
multipole moments. Namely, we have
\begin{equation}\label{PNmoment}
\int^1_{-1} dz~ \delta_\ell(z) \,\Sigma(\mathbf{x},t+z
\vert\mathbf{x}\vert/c) =
\sum_{k=0}^{+\infty}\,\frac{(2\ell+1)!!}{(2k)!!(2\ell+2k+1)!!}\,\left(\frac{
\vert\mathbf{x}\vert}{c}\frac{\partial}{\partial t}\right)^{2k}
\!\Sigma(\mathbf{x},t)\,.
\end{equation}
The above expression is then inserted into the right-hand side (RHS) of
Eqs.~\eqref{ILJL}, and truncated to fit with the PN order of the calculation.

A crucial finite part (FP) procedure is involved in the definition of
the source multipole moments~\eqref{ILJL}. It consists of (i)
multiplying the integrand of the moments by a regularization factor
$\vert\mathbf{x}\vert^B$ where $B$ is a complex
number,\,\footnote{Generally the regularization factor is taken to be
$\vert\mathbf{x}/r_0\vert^B$ where $r_0$ denotes some arbitrary constant
length scale. Here we set $r_0=1$ for convenience.} (ii) performing the
Laurent expansion when $B$ tends to the ``physical'' value $B=0$, and
(iii) picking up its finite part FP, namely the coefficient of the
zeroth power of $B$. The finite part regularization is therefore
equivalent to removing the poles $B^{-1}$, $B^{-2}$, ... (in the
analytically continued $B$-dependent integral) before taking the limit
$B\rightarrow 0$. The FP procedure is needed to compute the non-linear
contributions to the moments (generated by the gravitational source term
$\Lambda^{\mu\nu}$), which have a non-compact support extending up to
spatial infinity. Notice that no assumption nor physical restriction (in
principle) is involved, in the latter FP procedure, which has been
proved~\cite{B95,B96,B98mult} to yield the correct expression of the
multipole moments for general extended PN sources. It is precisely the
FP that guaranties this within the present formalism; the divergent
terms $B^{-1}$, $B^{-2}$, ... have no direct physical significance. Such
FP when $B\rightarrow 0$ is to be carefully distinguished from the
self-field regularization (\textit{e.g.}, Hadamard's or dimensional
regularization) which is to be invoked when treating the application of
the general formalism to singular point-particle sources.

We emphasize that the expressions~\eqref{ILJL} constitute \textit{a priori}
only a \textit{definition} of the source multipole moments. The point is that
such definition is fully related to the physical asymptotic wave form at
infinity from the source, which is computed using the multipolar
post-\textit{Minkowskian} formalism~\cite{BD86,B87}. In particular, the
\textit{radiative} multipole moments $U_L(t)$ and $V_L(t)$ which parameterize
the asymptotic wave form are given by some non-linear functionals of the
source moments~\eqref{ILJL}, made of many non-linear interactions between
them, including the famous GW tails corresponding to the interaction between
these moments and the total monopole mass $M$ of the source. The tails have
been computed within this approach in Refs.~\cite{BD92,B98tail}. However, for
deriving the spin terms at 1PN relative order, all these non-linear multipole
interactions, notably all the tails, are negligible (see further discussion
below). It is therefore sufficient to consider only the source multipole
moments of Eqs.~\eqref{ILJL}.

\section{Lowest-order spin effects in the multipole moments}\label{secIII}

The multipole moments discussed above can be applied to any source. Here, we
specialize them to binary systems of point-particles with spins. The
formalism to describe particles with spins was developed in
Refs.~\cite{Papa51spin,BOC75,BOC79,BI80}, and constitutes the basis of
most subsequent computations in this
field~\cite{KWW93,K95,MST96,TMSS96,OTO98,cho98,TOO01}. We reviewed this
formalism in paper~I and refer the reader to this paper for details
and notation. The spin contribution (marked by the underneath label S)
to the stress-energy tensor of the particles reads
\begin{equation}\label{Tmunu}
\mathop{T}_\text{S}{}^{\!\mu\nu}(t,\mathbf{x})=-\frac{1}{c}\,\nabla_
\rho\left[S_1^{\rho(\mu}\,v_1^{\nu)}
\frac{\delta(\mathbf{x}-\mathbf{y}_1)}{\sqrt{-g_1}}\right] +
1\leftrightarrow 2\,,
\end{equation}
where $\delta$ is the Dirac three-dimensional delta-function, $\nabla_\rho$
denotes the covariant derivative, $v_1^\mu(t)=(c,v_1^i)$ with
$v_1^i=dy_1^i/dt$ being the coordinate velocity of particle 1, $g_1$ is the
determinant of the metric evaluated at the location of particle 1 (following
Hadamard's self-field regularization), and $1\leftrightarrow 2$ means the same
expression as preceding, but for particle 2. The anti-symmetric spin tensor
$S_1^{\mu\nu}(t)$ is introduced in Sec.~II of paper~I. The covariant
four-vector $S^1_{\mu}$ is defined by
$S_1^{\mu\nu}=-\frac{1}{\sqrt{-g_1}}\varepsilon^{\mu\nu\rho\sigma}u^1_\rho
S^1_\sigma$; it is transverse to the particle's four-velocity,
$S^1_{\mu}u_1^{\mu}=0$. All results below are expressed in terms of some
particular space-like contravariant spin variables for each of the particles,
namely $S_1^i$ and $S_2^i$, the definition of which can be found in Eq.~(2.19)
of paper~I.

Similarly to the quantities $\Sigma_{\mu\nu}$ introduced in
Eq.~\eqref{Sigma}, we define the following matter-source densities,
depending on the components of the spin stress-energy
tensor~\eqref{Tmunu}:
\begin{subequations}\label{sigma}\begin{eqnarray}
&&\mathop{\sigma}_\text{S} \equiv
c^{-2}\left[\mathop{T}_\text{S}{}^{\!00}
+\mathop{T}_\text{S}{}^{\!ii}\right]~~\text{with
$\mathop{T}_\text{S}{}^{\!ii}=\delta_{ij}\mathop{T}_\text{S}{}^{\!ij}$}\,,\\
&&\mathop{\sigma}_\text{S}{}_{\!i} \equiv
c^{-1}\,\mathop{T}_\text{S}{}^{\!0i}\,,\\
&&\mathop{\sigma}_\text{S}{}_{\!ij} \equiv
\mathop{T}_\text{S}{}^{\!ij}\,.
\end{eqnarray}\end{subequations}
They are such that their ``non-spin'' counterparts (say
${}_\text{NS}\sigma_{\mu\nu}$) admit a finite non-zero limit when
$c\rightarrow +\infty$. They read:
\begin{subequations}\label{sigmaN}\begin{eqnarray}
\mathop{\sigma}_\text{S} &=&
-\frac{2}{c^3}\,\varepsilon_{ijk}\,v_1^i\,S_1^j\,\partial_k\delta_1+1\leftrightarrow
2+\mathcal{O}\left(\frac{1}{c^5}\right)\,,\\
\mathop{\sigma}_\text{S}{}_{\!i} &=&
-\frac{1}{2c}\,\varepsilon_{ijk}\,S_1^j\,\partial_k\delta_1+1\leftrightarrow
2+\mathcal{O}\left(\frac{1}{c^3}\right)\,,\\
\mathop{\sigma}_\text{S}{}_{\!ij} &=&
-\frac{1}{c}\,\varepsilon_{kl(i}\,v_1^{j)}\,S_1^k\,\partial_l\delta_1+1\leftrightarrow
2+\mathcal{O}\left(\frac{1}{c^3}\right)\,,
\end{eqnarray}\end{subequations}
where we denote $\delta_1\equiv\delta(\mathbf{x}-\mathbf{y}_1)$ and where
$\partial_k\delta_1$ means the gradient of $\delta_1$ with respect to the
field point $\mathbf{x}=(x^k)$. As shown by Eqs.~\eqref{sigmaN}, the leading
order of the vector and tensor densities ${}_\text{S}\sigma_i$ and
${}_\text{S}\sigma_{ij}$ is 0.5PN $\sim 1/c$. However, the scalar density
${}_\text{S}\sigma$ starts at a higher level, being at least 1.5PN $\sim
1/c^3$. At leading order in the spins, the $\Sigma_{\mu\nu}$'s, which depend
on the contributions of both matter and gravitational fields according to
Eqs.~\eqref{EE} and~\eqref{Sigma}, will reduce to their compact-support
material parts, namely the ${}_\text{S}\sigma_{\mu\nu}$'s given by
Eqs.~\eqref{sigmaN}. Indeed, the non-compact support gravitational part, whose
origin lies in the source term $\Lambda^{\mu\nu}$ present in the RHS of the
field equations~\eqref{EE}, always appears at a sub-dominant level, $1/c^2$
beyond the leading PN order. Hence,
\begin{subequations}\label{Sigmasigma}\begin{eqnarray}
\mathop{\Sigma}_\text{S} &=& \mathop{\sigma}_\text{S}
+\mathcal{O}\left(\frac{1}{c^5}\right)\,,\\
\mathop{\Sigma}_\text{S}{}_{\!i} &=& \mathop{\sigma}_\text{S}{}_{\!i}
+\mathcal{O}\left(\frac{1}{c^3}\right)\,,\\
\mathop{\Sigma}_\text{S}{}_{\!ij} &=& \mathop{\sigma}_\text{S}{}_{\!ij}
+\mathcal{O}\left(\frac{1}{c^3}\right)\,.
\end{eqnarray}\end{subequations}
The non-compact-support gravitational source terms play a role in our
computations at the next-to-leading order only (see Sec.~\ref{secIV}). We
conclude that the dominant contribution to the multipole moments~\eqref{ILJL}
due to the spins is given by
\begin{subequations}\label{ILJLN}\begin{eqnarray}
\mathop{I}_\text{S}{}_{\!L} &=& \int d^3\mathbf{x}\,\left\{
\hat{x}_L\,\mathop{\sigma}_\text{S}
-\frac{4(2\ell+1)}{c^2(\ell+1)(2\ell+3)} \,\hat{x}_{iL}
\,\mathop{\dot{\sigma}}_\text{S}{}_{\!i}\right\} +
\mathcal{O}\left(\frac{1}{c^5}\right)\,,\\ \mathop{J}_\text{S}{}_{\!L}
&=& \varepsilon_{ab\langle i_\ell} \int d^3
\mathbf{x}\,\hat{x}_{L-1\rangle a}
\,\mathop{\sigma}_\text{S}{}_{\!b} +
\mathcal{O}\left(\frac{1}{c^3}\right)\,.
\end{eqnarray}\end{subequations}
Thus, the dominant order is 1.5PN $\sim 1/c^3$ for spins in the
mass-type moments $I_L$, but only 0.5PN $\sim 1/c$ in the current-type
ones $J_L$. It is then evident (mathematically and physically) that the
spin part of the current moments always dominate over that of the mass
moments. We insert the explicit values~\eqref{sigmaN} into
Eqs.~\eqref{ILJLN}, integrate in a straightforward way (resorting to an
integration by parts and using the properties of the delta-function) and get
\begin{subequations}\label{ILJLNexpl}\begin{eqnarray}
\mathop{I}_\text{S}{}_{\!L} &=& \frac{2\ell}{c^3(\ell+1)} \left[\ell
\,v_1^i\,S_1^j\,\varepsilon_{ij\langle i_\ell}\,y_1^{L-1\rangle} -
(\ell-1)\,y_1^i\,S_1^j\,\varepsilon_{ij\langle
i_\ell}\,v_1^{i_{\ell-1}}\,y_1^{L-2\rangle}\right]\nonumber\\ &+&
1\leftrightarrow 2 + \mathcal{O}\left(\frac{1}{c^5}\right)\,,\\
\mathop{J}_\text{S}{}_{\!L} &=& \frac{\ell+1}{2c} \,y_1^{\langle
L-1}\,S_1^{i_\ell\rangle} + 1\leftrightarrow 2 +
\mathcal{O}\left(\frac{1}{c^3}\right)\,.
\end{eqnarray}\end{subequations}
It is worth to mention that in this calculation, limited to the lowest
PN order, the spins can be considered as constant since their time
variations, as given by the precessional equations (see paper~I), are
always smaller by a factor $1/c^2$ at least.

\section{Higher-order spin effects in the multipole moments}\label{secIV}

For the present purpose, we need the spin contributions to the mass-quadrupole
moment $I_{ij}$ and the current-quadrupole moment $J_{ij}$ one PN order beyond
the leading terms obtained in Eqs.~\eqref{ILJLNexpl}. This means at 2.5PN
order for $I_{ij}$ and at 1.5PN order for $J_{ij}$. As said previously, the
non-linear gravitational source terms, with non-compact support, start playing
a role at the 1PN relative order.\,\footnote{The rule admits some exceptions,
  though. For instance, in the non-spinning part of the mass multipole moments
  $I_L$, one may have expected the non-linear non-compact gravitational terms
  to appear at 1PN order, but these terms turn out to be in the form of a pure
  divergence and can be integrated out to zero. As a result, the non-compact
  terms of the non-spinning parts contribute to $I_L$ at 2PN order only (and
  at 1PN order to $J_L$). See Refs.~\cite{BD89,B95} for details.} Therefore,
they do make a net contribution to the spin parts of both $I_{ij}$ at 2.5PN
order and $J_{ij}$ at 1.5PN order. We can note here that the authors of
Ref.~\cite{OTO98} computed $J_{ij}$ at 1.5PN order \textit{but} neglected all
the non-compact source terms in their calculation, thus obtaining an incorrect
result.

We now reduce the multipole moments to the required PN order by
inserting the expansion formula~\eqref{PNmoment} into the general
expressions~\eqref{ILJL}. Neglecting PN terms of higher
order, as indicated, we have
\begin{subequations}\label{ILJLpn}\begin{align}
&I_{ij} =\mathop{\mathrm{FP}}_{B=0}\,\int
d^3\mathbf{x}\,\vert\mathbf{x}\vert^B \left\{ \hat{x}_{ij}\,\Sigma +
\frac{1}{14c^2}\hat{x}_{ij}\vert\mathbf{x}\vert^2 \,\ddot{\Sigma} +
\frac{1}{504c^4}\hat{x}_{ij}\,\vert\mathbf{x}\vert^4
\,\ddddot{\Sigma}\right.\nonumber\\&\qquad\left.
-\frac{20}{21c^2}\,\hat{x}_{ijk} \,\dot{\Sigma}_k - \frac{10}{189
c^4}\,\hat{x}_{ijk}\,\vert\mathbf{x}\vert^2\,\dddot{\Sigma}_k
+\frac{5}{54c^4}\,\hat{x}_{ijkl}\ddot{\Sigma}_{kl}\right\} +
\mathcal{O}\left(\frac{1}{c^6}\right)\,,\\ &J_{ij} =
\mathop{\mathrm{FP}}_{B=0}\,\varepsilon_{ab\langle i} \int d^3
\mathbf{x}\,\vert\mathbf{x}\vert^B \biggl\{ \hat{x}_{j\rangle a}
\,\Sigma_b + \frac{1}{14c^2}\hat{x}_{j\rangle a}\,\vert\mathbf{x}\vert^2
\,\ddot{\Sigma}_b -\frac{5}{28c^2}\,\hat{x}_{j\rangle ac}
\,\dot{\Sigma}_{bc}\biggr\} + \mathcal{O}\left(\frac{1}{c^4}\right)\,,
\end{align}\end{subequations}
where FP denotes the essential process of extracting the finite
part, as explained in Sec.~\ref{secII}. We can further reduce
Eqs.~\eqref{ILJLpn} by substituting the appropriate expressions of the
source terms $\Sigma_{\mu\nu}$ as given by Eqs.~(4.11) of
Ref.~\cite{B95}. For completeness, we list all the necessary expressions below:
\begin{subequations}\label{Sigmamunu}\begin{align}
 \Sigma &= \left[ 1+\frac{4V}{c^2} +\frac{2}{c^4} (\hat{W}+4V^2)\right]
 \sigma - \frac{1}{\pi Gc^2}\,\partial_i V \partial_i V \nonumber \\ & +
 \frac{1}{\pi Gc^4} \biggl\{ -V \partial^2_t V- 2V_i\partial_t
 \partial_i V - \hat{W}_{ij}\partial_{ij} V-\frac{1}{2} (\partial_t
 V)^2 \nonumber \\ & \qquad +2 \partial_i V_j \partial_j V_i -
 \partial_i V\partial_i \hat{W} - \frac{7}{2} V\partial_i V\partial_i V
 \biggr\} + \mathcal{O}\left(\frac{1}{c^6}\right)\,,\label{Sigma0}\\
 \Sigma_i &= \left[ 1 +\frac{4V}{c^2}\right] \sigma_i + \frac{1}{\pi
 Gc^2} \left\{ \partial_k V(\partial_i V_k -\partial_k V_i) +
 \frac{3}{4} \partial_t V \partial_i V \right\} +
 \mathcal{O}\left(\frac{1}{c^4}\right)\,,\\ \Sigma_{ij} &= \sigma_{ij} +
 \frac{1}{4\pi G} \left\{ \partial_i V \partial_j V - \frac{1}{2}
 \delta_{ij} \partial_k V \partial_k V\right\} +
 \mathcal{O}\left(\frac{1}{c^2}\right)\,,
\end{align}\end{subequations}
where the material source densities are given by $\sigma =
c^{-2}\left[T^{00} +T^{ii}\right]$, $\sigma_i = c^{-1}\,T^{0i}$ and
$\sigma_{ij} = T^{ij}$. The non-compact support terms in
Eqs.~\eqref{Sigmamunu} are parameterized by a particular set of
``elementary'' potentials $V$, $V_i$, $\hat{W}_{ij}$, ..., which enter
the harmonic-coordinate near zone metric at the 2PN order computed in
Ref.~\cite{BFP98}.\,\footnote{The potential called $W_{ij}$ in
Ref.~\cite{B95} differs from the present $\hat{W}_{ij}$ (whose
definition is given in paper~I) according to the formula
$W_{ij}=\hat{W}_{ij}-\frac{1}{2}\delta_{ij}\hat{W}$, hence
$W=W_{ii}=-\frac{1}{2}\hat{W}$.} Their complete expressions are given in
Sec.~III of paper~I.

We can simplify $\Sigma$ substantially by using some identities of the
type $\partial_iA\partial_iB=\frac{1}{2}\left[\Delta\left(A
B\right)-A\Delta B-B\Delta A\right]$, replacing the Laplacians $\Delta
A$ and $\Delta B$ by their PN sources, and disregarding the pure
Laplacian term $\frac{1}{2}\Delta\left(A B\right)$ because it makes zero
contribution to the moment. This last point comes essentially from the
fact that, after integration by parts, a pure Laplacian term in the moment
sources is equivalent to a source term proportional to
some $\Delta \hat{x}_L = 0$. Beware that, because of the presence of the
finite part FP, such a ``proof'' is not correct in general and, in fact,
the pure Laplacian terms do generally contribute at high PN orders.
However, for the terms under concern, merely at the 1PN relative order,
the argument can be made rigorous and shown to work, so that we can
indeed discard these pure Laplacians in the present computation (see
Ref.~\cite{B95} for the proof). Hence, we get the simpler formula,
\begin{eqnarray}\label{Sigma'}
 \Sigma' &=& \sigma + \frac{4V}{c^4}\sigma_{ii} + \frac{1}{\pi Gc^4}
 \left\{ - 2V_i\partial_t \partial_i V -
 \hat{W}_{ij}\partial_{ij} V-\frac{1}{2} (\partial_t V)^2 + 2
 \partial_i V_j \partial_j V_i \right\} \nonumber\\&+&
 \mathcal{O}\left(\frac{1}{c^6}\right)\,,
\end{eqnarray}
which differs from $\Sigma$ by pure Laplacian terms of the type
$\sim\Delta\left(A B\right)$. We have checked that the two different
forms $\Sigma$ and $\Sigma'$, Eqs.~\eqref{Sigma0} and~\eqref{Sigma'},
lead to the same final result.

\subsection{Compact--support contribution}

Having now the general set up for our computation we consider first the
compact-support part of the multipole moments, \textit{i.e.} that part
proportional to the material source densities $\sigma_{\mu\nu}$, and
given by the first terms in Eqs.~\eqref{Sigmamunu}--\eqref{Sigma'}. For
these terms we make two computations. The first one consists of (i)
evaluating all the components of $T^{\mu\nu}$ to the correct
PN order [extending thus Eqs.~\eqref{sigmaN} and including the monopolar
sources, which may give rise to spin terms through the metric], (ii) computing
their time derivatives by making use of the usual replacement of accelerations
by the equations of motion, and of the time-derivatives of the spins by the
precessional equations, (iii) transforming the time-derivatives, when
applied to delta-functions, into spatial ones using the formula
$\partial_t\delta_1=-v_1^i\partial_i\delta_1$, (iv) operating by parts
the spatial derivatives of delta-functions to finally integrate thanks
to the basic property of delta-functions.

Normally, such basic property of delta-functions reads $\int d^3
\mathbf{x} \,F(\mathbf{x}) \, \delta_1(\mathbf{x}) = (F)_1$, where
$(F)_1$ is simply the value of the function at the point 1. When the
function is regular, there is no problem and we have
$(F)_1=F(\mathbf{y}_1,t)$, for instance $(\hat{x}_L)_1=\hat{y}_1^L$.
However, when the function $F$ is singular at the point 1 (\textit{i.e.}
when $\mathbf{x}\rightarrow\mathbf{y}_1$), a choice must be made for a
``self-field'' regularization, able to subtract the infinities in a
consistent way. There are various possibilities. In the present work we
adopt, following paper~I, the Hadamard self-field regularization. At the
order we are working (relative 1PN order) the various possible choices
are equivalent. For instance, one can show that dimensional
regularization would give the same result as Hadamard's regularization,
essentially because at such low PN order there are no poles in the
dimension of space [say, $\propto (d-3)^{-1}$], which correspond to
logarithmic divergences in Hadamard's regularization. We then define
$(F)_1$ to be given by the \textit{partie finie} of the function $F$ at
point 1 in the sense of Hadamard (see \textit{e.g.}~\cite{BFreg} for a
full account of this regularization). Suppose for example that
$F=U\,\hat{x}_L$ where $U=\frac{G m_1}{r_1}+\frac{G m_2}{r_2}$ is the
Newtonian potential of point particles, singular at the location of the
particles. Then we easily compute that Hadamard's partie finie is
$(F)_1=\frac{G m_2}{r_{12}}\,\hat{y}_1^L$.

Our alternative computation of the compact-support terms in the moments
is the same as the one performed by Owen \textit{et al.}~\cite{OTO98}.
It consists of applying the following formula (derived in~\cite{OTO98}),
\begin{align}\label{OTO}
\int
d^3\mathbf{x}\,F(\mathbf{x},t)\,\mathop{T}_\text{S}{}^{\!\mu\nu}(t,\mathbf{x})
&= -\frac{d}{c
dt}\left[S_1^{0(\mu}\,v_1^{\nu)}\,\frac{(F)_1}{\sqrt{-g_1}}\right]+S_1^{\rho(\mu}\,
v_1^{\nu)} \,\frac{(\partial_\rho F)_1}{\sqrt{-g_1}}\nonumber\\ &+
\left[\mathop{\Gamma}_1{}^{\!(\mu}_{\!\rho\sigma}\,S_1^{\nu)\rho}\,v_1^{\sigma}-
\mathop{\Gamma}_1{}^{\!\rho}_{\!\sigma\rho}\,S_1^{\sigma(\mu}\,v_1^{\nu)}\right]
\frac{(F)_1}{\sqrt{-g_1}} + 1\leftrightarrow 2\,,
\end{align}
which is valid for any function $F(\mathbf{x},t)$, and where $(F)_1$ and
$(\partial_\rho F)_1$ have to be understood as the Hadamard partie finie of
the function and its derivative. This formula is very useful but must be
handled with care. In particular, when computing $(\partial_\rho F)_1$ in the
second term of the RHS of~\eqref{OTO}, we notice that the gradient is to be
taken \textit{first}, and only then should one deduce the value at point 1.
The result can be different if one permutes the order of operations. Suppose
for instance that one is computing $(\partial_t F)_1$ where $F=\hat{x}_L$.
Clearly, since $\partial_t\,\hat{x}_L=0$ the result is zero. However, if one
computes $\frac{d}{d t}(F)_1$ instead of $(\partial_t F)_1$, one obtains an
incorrect non-zero result, which is equal in this case to $\frac{d}{d
  t}\!\left[\hat{y}_1^L\right] = \ell y_1^{\langle L-1}v_1^{i_\ell\rangle}$.
We found that this error, \textit{i.e.} computing $\frac{d}{d t} (\hat{x}_L)_1
= \ell y_1^{\langle L-1}v_1^{i_\ell\rangle}$ instead of
$(\partial_t\,\hat{x}_L)_1=0$, was committed in the evaluation of the
(compact-support part of the) current quadrupole moment $J_{ij}$ in
Ref.~\cite{OTO98}.

\subsection{Non-compact--support contribution}

We now derive the non-compact support part of the multipole moments.
Inspection of the expressions~\eqref{Sigmamunu}--\eqref{Sigma'} shows
that we need only the elementary potentials $V$, $V_i$, $\hat{W}_{ij}$
(and $\hat{W}\equiv\hat{W}_{ii}$) at their \textit{lowest} PN order in
the spins. They read:
\begin{subequations}\label{potS}\begin{align}
\mathop{V}_\text{S} &= -
\frac{2G}{c^3}\,\varepsilon_{ijk}\,v_1^i\,S_1^j\,\partial_k\left(\frac{
1}{r_1}\right)+1\leftrightarrow
2+\mathcal{O}\left(\frac{1}{c^{5}}\right)\,,\\
\mathop{V}_\text{S}{}_{\!i} &= - \frac{G}{2
c}\,\varepsilon_{ijk}\,S_1^j\,\partial_k\left(\frac{1}{r_1}
\right)+1\leftrightarrow 2+\mathcal{O}\left(\frac{1}{c^{3}}\right)\,,\\
\mathop{\hat{W}}_\text{S}{}_{\!ij} &= - \frac{G}{c}
\,\varepsilon_{kl(i}\,v_1^{j)}\,S_1^k\,\partial_l\left(\frac{1}{r_1}\right)
+
\frac{G}{c}\,\delta_{ij}\,\varepsilon_{klm}\,v_1^k\,S_1^l\,\partial_m\left(\frac{1}
{r_1}\right)+1\leftrightarrow
2+\mathcal{O}\left(\frac{1}{c^{3}}\right)\,,\\ \mathop{\hat{W}}_\text{S}
&= \frac{2G}{c}
\,\varepsilon_{klm}\,v_1^k\,S_1^l\,\partial_m\left(\frac{1}{r_1}\right)
+1\leftrightarrow 2+\mathcal{O}\left(\frac{1}{c^{3}}\right)\,.
\end{align}\end{subequations}
The latter spin contributions enter only the ``non-spin'' parts of
Eqs.~\eqref{Sigmamunu}--\eqref{Sigma'}. We need now to perform the
integrations in Eq.~\eqref{ILJLpn}. The calculation becomes simple if we
use the following trick. In the multipole moment's integrands we
transform all the gradients evaluated at the field point $\mathbf{x}$
into gradients evaluated at the source points $\mathbf{y}_1$ or
$\mathbf{y}_2$ using \textit{e.g.} $\frac{\partial}{\partial
x^i}\left(1/r_1\right)=-\frac{\partial}{\partial
y_1^i}\left(1/r_1\right)$. Then, we put the source-type gradients
$\frac{\partial}{\partial y_1^i}$ and $\frac{\partial}{\partial y_2^i}$
outside the integrals and we express the result solely in terms of the
function
\begin{subequations}\begin{equation}\label{YL}
Y_L(\mathbf{y}_1,\mathbf{y}_2)\equiv
-\frac{1}{2\pi}\mathop{\mathrm{FP}}_{B=0}\,\int
d^3\mathbf{x}\,\vert\mathbf{x}\vert^B \frac{\hat{x}_{L}}{r_1 r_2}\,,
\end{equation}
which is known to admit the analytically closed-form~\cite{B95}
\begin{equation}\label{YLcf}
Y_L = \frac{r_{12}}{\ell+1}\sum_{p=0}^\ell y_1^{\langle
L-P}y_2^{P\rangle}\,.
\end{equation}\end{subequations}
Thus, the closed-form expressions of the non-compact (NC)
parts of the spin multipole moments (they depend on the function $Y_L$
for $\ell=2,3$) are:
\begin{subequations}\begin{align}
\mathop{I}_\text{S}{}_{\!ij}^{\mathrm{(NC)}} &=\frac{2G
m_2}{c^5}\left\{\varepsilon_{mnk}\,v_1^l\,S_1^m\,\mathop{\partial}_{1}{}_{\!n}
\mathop{\partial}_{2}{}_{\!kl}Y_{ij}-
\varepsilon_{mnp}\,v_1^m\,S_1^n\,\mathop{\partial}_{1}{}_{\!p}
\mathop{\Delta}_{2}Y_{ij}\right.\nonumber\\ &\qquad -
\varepsilon_{kmn}\,v_2^l\,S_1^m\,\mathop{\partial}_{1}{}_{\!n}
\mathop{\partial}_{2}{}_{\!kl}Y_{ij} -
2\varepsilon_{lmn}\,v_2^k\,S_1^m\,\mathop{\partial}_{1}{}_{\!kn}
\mathop{\partial}_{2}{}_{\!l}Y_{ij}\nonumber\\ &\qquad\left.
+\frac{10}{21}\,\varepsilon_{lmn}\,S_1^m\,\frac{d}{dt}\left[
\mathop{\partial}_{1}{}_{\!nk}\mathop{\partial}_{2}{}_{\!l}Y_{ijk}\right]
-\frac{10}{21}\,\varepsilon_{kmn}\,S_1^m\,\frac{d}{dt}\left[
\mathop{\partial}_{1}{}_{\!nl}\mathop{\partial}_{2}{}_{\!l}Y_{ijk}
\right]\right\}\nonumber\\&+1\leftrightarrow 2 +
\mathcal{O}\left(\frac{1}{c^7}\right)\,,\label{IijSnc}\\
\mathop{J}_\text{S}{}_{\!ij}^{\mathrm{(NC)}} &=\frac{G
m_2}{c^3}\varepsilon_{kl\langle
i}\left\{-\varepsilon_{kmn}\,S_1^m\,\mathop{\partial}_{2}{}_{\!k}
\mathop{\partial}_{1}{}_{\!ln}Y_{j\rangle
k}+\varepsilon_{lmn}\,S_1^m\,\mathop{\partial}_{2}{}_{\!p}
\mathop{\partial}_{1}{}_{\!pn}Y_{j\rangle
k}\right\}\nonumber\\&+1\leftrightarrow 2 +
\mathcal{O}\left(\frac{1}{c^5}\right)\,,\label{JijSnc}
\end{align}\end{subequations}
where $\partial_{1k}\equiv\partial/\partial y_1^k$ and
$\partial_{2k}\equiv\partial/\partial y_2^k$. The final computation of
the moments using formula~\eqref{YLcf} is straightforward.

\section{Results for the multipole moments and flux}\label{secV}

The expressions for the multipole moments ${}_\text{S}I_{ij}$ and
${}_\text{S}J_{ij}$, including both compact and non-compact
contributions as computed in Sec.~\ref{secIV}, are quite long if written
in a general frame. They can be substantially simplified by going to the
frame of the center-of-mass (CM). When working in the CM frame it is
convenient to use the following spin variables:
\begin{subequations}\label{SSigma}\begin{align}
\mathbf{S} &\equiv \mathbf{S}_1 + \mathbf{S}_2\,,\\ \mathbf{\Sigma}
&\equiv m\left[\frac{\mathbf{S}_2}{m_2} -
\frac{\mathbf{S}_1}{m_1}\right]\,.
\end{align}\end{subequations}
These spin variables were initially introduced by Kidder~\cite{K95} except that
here we denote $\mathbf{\Sigma}$ what he calls $\mathbf{\Delta}$.
Mass parameters will be denoted by $m\equiv m_1+m_2$, $\delta m \equiv
m_1-m_2$ and $\nu\equiv m_1\,m_2/m^2$ for a mass ratio such that
$\nu=1/4$ for equal masses and $\nu\rightarrow 0$ in the test-mass
limit.

The CM frame is defined by the nullity of the binary's dipole moment or
equivalently the CM vector $\mathbf{G}$. At 2.5PN order including spin
effects, it can easily be determined using the vector $\mathbf{G}$
evaluated in paper~I. However, here we need only the lowest order term
(1.5PN in the spins) together with the 1PN non-spin correction; the
2.5PN term in the spins cancels out. To the needed order we have (see
\textit{e.g.} Eq.~(5.13) in~\cite{TOO01})
\begin{subequations}\label{CMpos}\begin{align}
\mathbf{y}_1 &= \left[\frac{m_2}{m} + \frac{\nu}{2 c^2}\frac{\delta
m}{m}\left(v^2-\frac{G\,m}{r}\right)\right]\,\mathbf{x} +
\frac{\nu}{m\,c^3}\,\mathbf{v}\times\mathbf{\Sigma}\,,\\ \mathbf{y}_2 &=
\left[-\frac{m_1}{m} + \frac{\nu}{2 c^2}\frac{\delta
m}{m}\left(v^2-\frac{G\,m}{r}\right)\right]\,\mathbf{x} +
\frac{\nu}{m\,c^3}\,\mathbf{v}\times\mathbf{\Sigma}\,,
\end{align}\end{subequations}
which gives the CM positions of the particles, $\mathbf{y}_1$ and
$\mathbf{y}_2$, in terms of the relative position and velocity,
$\mathbf{x}=\mathbf{y}_1-\mathbf{y}_2$ and
$\mathbf{v}=d\mathbf{x}/dt=\mathbf{v}_1-\mathbf{v}_2$ (we pose
$r=\vert\mathbf{x}\vert$ and $v^2=\mathbf{v}\cdot\mathbf{v}$).

\subsection{The multipole moments}

Our final result for the spin part of the mass-quadrupole moment at
2.5PN order (1PN order beyond the dominant SO term), for general orbits
and in the CM frame, reads
\begin{eqnarray}\label{IijS}
\mathop{I}_\text{S}{}_{\!ij} &=& \frac{\nu}{c^3}\biggl\{\frac{8}{3}
\,x^{\langle i}(\mathbf{v}\times\mathbf{S})^{j\rangle} - \frac{4}{3}
\,v^{\langle i}(\mathbf{x}\times\mathbf{S})^{j\rangle}\nonumber\\
&&\quad + \frac{8}{3}\,\frac{\delta m}{m} \,x^{\langle
i}(\mathbf{v}\times\mathbf{\Sigma})^{j\rangle} -
\frac{4}{3}\,\frac{\delta m}{m} \,v^{\langle
i}(\mathbf{x}\times\mathbf{\Sigma})^{j\rangle}\biggr\}\nonumber\\
&+&\frac{\nu}{c^5}\biggl\{\left[\frac{5}{3}+\frac{2}{7}\nu\right]\frac{G
m}{r^3}\,\frac{\delta m}{m} \,(x v) \,x^{\langle
i}(\mathbf{x}\times\mathbf{\Sigma})^{j\rangle}\nonumber\\ &&\quad +
\left(\left[\frac{7}{3}+4 \nu\right]\frac{G m}{r} +
\left[\frac{26}{21}-\frac{116}{21}\nu\right] v^2\right)\frac{\delta m}{m}
\,x^{\langle i}(\mathbf{v}\times\mathbf{\Sigma})^{j\rangle}\nonumber\\
&&\quad + \left[\frac{31}{21}+\frac{19}{21}\nu\right]\frac{G
m}{r^3}\,(x v) \,x^{\langle
i}(\mathbf{x}\times\mathbf{S})^{j\rangle}\nonumber\\ &&\quad +
\left(\left[\frac{25}{7}+\frac{55}{21}\nu\right]\frac{G m}{r} +
\left[\frac{26}{21}-\frac{26}{7}\nu\right] v^2\right) x^{\langle
i}(\mathbf{v}\times\mathbf{S})^{j\rangle}\nonumber\\ &&\quad +
\left(\left[-4 -\frac{2}{7}\nu\right]\frac{G m}{r} +
\left[-\frac{6}{7}+\frac{64}{21}\nu\right] v^2\right) \frac{\delta m}{m}
\,v^{\langle i}(\mathbf{x}\times\mathbf{\Sigma})^{j\rangle}\nonumber\\
&&\quad + \left[\frac{10}{21}-\frac{8}{21}\nu\right]\frac{\delta m}{m}
\,(x v) \,v^{\langle
i}(\mathbf{v}\times\mathbf{\Sigma})^{j\rangle}\nonumber\\ &&\quad +
\left(\left[-\frac{26}{3}-\frac{2}{3}\nu\right]\frac{G m}{r} +
\left[-\frac{6}{7}+\frac{18}{7}\nu\right] v^2\right) v^{\langle
i}(\mathbf{x}\times\mathbf{S})^{j\rangle}\nonumber\\ &&\quad +
\left[\frac{10}{21}-\frac{10}{7}\nu\right] (x v) \,v^{\langle
i}(\mathbf{v}\times\mathbf{S})^{j\rangle}\nonumber\\ &&\quad +
\left(\left[\frac{52}{21}-\frac{10}{7}\nu\right](S,x,v)+
\left[\frac{62}{21}-\frac{18}{7}\nu\right]\frac{\delta
m}{m}\,(\Sigma,x,v)\right)\frac{G m}{r^3} \,x^{\langle
i}x^{j\rangle}\nonumber\\ &&\quad +
\left(\left[-\frac{5}{21}+\frac{5}{7}\nu\right](S,x,v)+
\left[-\frac{5}{21}-\frac{4}{7}\nu\right]\frac{\delta
m}{m}\,(\Sigma,x,v)\right)v^{\langle i}v^{j\rangle} \nonumber\\ &&\quad
+ \left(\left[-\frac{8}{3}+\frac{16}{3}\nu\right](x S)+
\left[-\frac{8}{3}+\frac{8}{3}\nu\right]\frac{\delta m}{m}\,(x
\Sigma)\right)\frac{G m}{r^3} x^{\langle
i}(\mathbf{x}\times\mathbf{v})^{j\rangle}\nonumber\\ &&\quad +
\left(\left[\frac{4}{3}-4\nu\right](v S)+
\left[\frac{4}{3}-\frac{8}{3}\nu\right]\frac{\delta m}{m}\,(v
\Sigma)\right)v^{\langle
i}(\mathbf{x}\times\mathbf{v})^{j\rangle}\biggr\}\nonumber\\&+&
\mathcal{O}\left(\frac{1}{c^7}\right)\,.
\end{eqnarray}
The scalar product of ordinary Euclidean vectors is indicated by
parenthesis, \textit{e.g.} $(v S)=\mathbf{v}\cdot\mathbf{S}$, the
cross product by the usual cross symbol,
$(\mathbf{x}\times\mathbf{\Sigma})^i=\varepsilon^{ijk}x^j\Sigma^k$, and
the mixed product of three vectors by $(S,x,v)=\mathbf{S}\cdot
(\mathbf{x}\times\mathbf{v})=\varepsilon^{ijk}S^ix^jv^k$. We recall also
that the STF projection is denoted using carets surrounding indices, 
\textit{i.e.} $\langle ij\rangle$. Next, the spin part of the current quadrupole
moment at 1.5PN order (also 1PN order beyond the leading term) is
\begin{eqnarray}\label{JijS}
\mathop{J}_\text{S}{}_{\!ij} &=&
\frac{\nu}{c}\biggl\{-\frac{3}{2}\,x^{\langle
i}\Sigma^{j\rangle}\biggr\}\nonumber\\ &+&
\frac{\nu}{c^3}\biggl\{\left[\frac{3}{7}-\frac{16}{7}\nu\right] (x v)
\,v^{\langle i}\Sigma^{j\rangle} + \frac{3}{7} \,\frac{\delta m}{m} \,(x
v) \,v^{\langle i}\,S^{j\rangle}\nonumber\\&&\quad +
\left(\left[\frac{27}{14}-\frac{109}{14}\nu\right] (v \Sigma) +
\frac{27}{14} \frac{\delta m}{m} \,(v S)\right) x^{\langle
i}v^{j\rangle}\nonumber\\&&\quad +
\left(\left[-\frac{11}{14}+\frac{47}{14}\nu\right] (x \Sigma) -
\frac{11}{14} \frac{\delta m}{m} \,(x S)\right) v^{\langle
i}v^{j\rangle}\nonumber\\&&\quad +
\left(\left[\frac{19}{28}+\frac{13}{28}\nu\right] \frac{G m}{r} +
\left[-\frac{29}{28}+\frac{143}{28}\nu\right] v^2\right) x^{\langle
i}\Sigma^{j\rangle}\nonumber\\&&\quad +
\left(\left[-\frac{4}{7}+\frac{31}{14}\nu\right] (x \Sigma) -
\frac{29}{14} \frac{\delta m}{m} \,(x S)\right) \frac{G m}{r^3}
\,x^{\langle i}x^{j\rangle}\nonumber\\&&\quad +
\left(-\frac{1}{14}\frac{G m}{r} -\frac{2}{7} v^2\right) \frac{\delta
m}{m} \,x^{\langle i}S^{j\rangle}\biggr\}\nonumber\\ &+&
\mathcal{O}\left(\frac{1}{c^5}\right)\,.
\end{eqnarray}
Notice that the 1.5PN current quadrupole moment ${}_\text{S}J_{ij}$ was
also computed in Ref.~\cite{OTO98}, see Eq.~(4.18) there. However our
result~\eqref{JijS} differs from their result. There are two reasons for
this discrepancy. The main reason is that Ref.~\cite{OTO98} completely
neglected the non-compact support terms, which originate from the
non-linearities of the Einstein field equations \textit{via} the term
$\Lambda^{\mu\nu}$ in Eq.~\eqref{EE}, and physically represent the
gravitational field acting as a source for the multipole moment. As we
have seen in Sec.~\ref{secIV} these terms are not negligible. Their
contribution to the 1.5PN order current moment ${}_\text{S}J_{ij}$ has
been computed in Eq.~\eqref{JijSnc}. The second reason for the
difference between Eq.~\eqref{JijS} and the result of \cite{OTO98} is a
computational error in~\cite{OTO98} when they apply the integration
formula~\eqref{OTO} for computing the compact-support terms. We have
already commented upon this error after Eq.~\eqref{OTO} above. These two
errors fully account for the discrepancy between our result and the one
of Eq.~(4.18) in Ref.~\cite{OTO98}.

Next, in order to derive the GW flux at 2.5PN order, we need the spin
parts in the mass octupole and current octupole moments, but only at the
lowest order in the spins. They can be obtained from our previous
computation leading to Eqs.~\eqref{ILJLNexpl}, after the CM reduction.
We find\,\footnote{The result for ${}_\text{S}I_{ijk}$ agrees with the
one given by Eq.~(4.17) in Ref.~\cite{OTO98}. However, we notice a
misprint in the second term of Eq.~(4.17) in~\cite{OTO98}, in which
$x^{jk}$ should read $x^jv^k$.}
\begin{subequations}\label{octIJ}\begin{eqnarray}
\mathop{I}_\text{S}{}_{\!ijk} &=&
\frac{\nu}{c^3}\biggl\{-\frac{9}{2}\,\frac{\delta m}{m} \,x^{\langle
i}x^j(\mathbf{v}\times\mathbf{S})^{k\rangle}-\frac{3}{2}\,(3-11\nu)\,x^{\langle
i}x^j(\mathbf{v}\times\mathbf{\Sigma})^{k\rangle}\nonumber\\ &&\quad
+3\,\frac{\delta m}{m} \,x^{\langle
i}v^j(\mathbf{x}\times\mathbf{S})^{k\rangle}+3\,(1-3\nu)\,x^{\langle
i}v^j(\mathbf{x}\times\mathbf{\Sigma})^{k\rangle}\biggr\}+
\mathcal{O}\left(\frac{1}{c^5}\right)\,,\\ \mathop{J}_\text{S}{}_{\!ijk}
&=&\frac{\nu}{c}\biggl\{2\,x^{\langle i}x^jS^{k\rangle}+2\,\frac{\delta
m}{m} \,x^{\langle i}x^j\Sigma^{k\rangle}\biggr\} +
\mathcal{O}\left(\frac{1}{c^3}\right)\,.
\end{eqnarray}\end{subequations}

\subsection{The gravitational-wave energy flux}

With all these moments, Eqs.~\eqref{IijS}--\eqref{JijS}
and~\eqref{octIJ}, and only with those, we can compute the 2.5PN spin
part of the GW flux. Indeed, recall that the spins start at 1.5PN order
in the mass moments and at 0.5PN order in the current ones, so one can
easily see that in higher multipoles spins will enter the flux at higher PN
order. On the other hand, one can check that it is not necessary to
include the effects of tails of GWs, and more generally of any
non-linear multipole interaction. Indeed, the tails give a correction to
each of the source-type multipole moments $I_L$ and $J_L$ at the
relative order 1.5PN $\sim 1/c^3$ (see \textit{e.g.}~\cite{B95}). For
the mass quadrupole $I_{ij}$ the spin itself is at order 1.5PN so the
tail will arise only at order 3PN $\sim 1/c^6$ in the flux. For the
current quadrupole $J_{ij}$ the spin is at 0.5PN but $J_{ij}$ comes in
the flux at 1PN order, so again we see that the corresponding tail will
only be at 3PN order in the flux. In conclusion, for this problem it is
sufficient to express the flux solely in terms of the source multipole
moments $I_L$ and $J_L$; all multipole interactions built in the
radiative moments seen at infinity, namely $U_L$ and $V_L$, are
negligible. Furthermore, as we have seen only four multipolar
contributions are important for this application. Therefore (\textit{cf.}
Eq.~(4.28) in~\cite{B95})
\begin{eqnarray}\label{flux}
\mathcal{F} &=&
\frac{G}{c^5}\left\{\frac{1}{5}\dddot{I}_{ij}\dddot{I}_{ij}+\frac{1}{c^2}\left[
\frac{1}{189}\ddddot{I}_{ijk}\ddddot{I}_{ijk}
+\frac{16}{45}\dddot{J}_{ij}\dddot{J}_{ij}\right]
+\frac{1}{84c^4}\ddddot{J}_{ijk}\ddddot{J}_{ijk}\right\}\\
&+&\text{terms not contributing to the spins at 2.5PN order}\,.\nonumber
\end{eqnarray}
In order to compute time derivatives of the
moments~\eqref{IijS}--\eqref{JijS} and~\eqref{octIJ}, we must be careful
at including the non-spin terms of $I_{ij}$ and $J_{ij}$ since these
terms will generate by order reduction of the accelerations some new
spin terms at 2.5PN order. In particular we need for this problem the
non-spin part of the mass quadrupole $I_{ij}$ at 1PN order, which is
given by~\cite{BS89,BI04mult},
\begin{align}\label{quad1PN}
I_{ij} &= m\,\nu\left\{\left(1+\frac{1}{c^2}\left[\left(\frac{29}{42}
  -\frac{29}{14}\nu\right)v^2+\left(-\frac{5}{7}
  +\frac{8}{7}\nu\right)\frac{G\,m}{r}\right]\right) x^{\langle
  i}x^{j\rangle} \right.\nonumber\\&\qquad\quad\left. +
\frac{r^2}{c^2}\left(\frac{11}{21} -\frac{11}{7}\nu\right) v^{\langle
  i}v^{j\rangle} + \frac{(xv)}{c^2}\,\left (
-\frac{4}{7}+\frac{12}{7}\,\nu \right )\, x^{\langle i}\,v^{j
  \rangle}\right\} + \mathcal{O}\left(\frac{1}{c^4}\right)\,.
\end{align}
The current quadrupole moment is also needed at 1PN order,
\begin{align}\label{quadcurr1PN}
J_{ij} &= -\nu\,\delta m \,x^{k}v^{l}\,\varepsilon^{kl\langle i}\left\{
x^{j\rangle} \left(1 + \frac{1}{c^2}\left[\left
(\frac{27}{14}+\frac{15}{7}\, \nu \right ) \frac{G m}{r}+ \left (
\frac{13}{28}-\frac{17}{7}\nu \right ) v^2 \right]\right)
\right.\nonumber \\ & \qquad\qquad\left. + \frac{1}{c^2}v ^{j\rangle}
(xv) \left (\frac{5}{28}-\frac{5}{14}\, \nu \right )\right\} +
\mathcal{O}\left(\frac{1}{c^{4}}\right)\,.
\end{align}
Note that for both $I_{ij}$ and $J_{ij}$ there are some contributions at
1.5PN order which depends on the spin variables and are generated from
the Newtonian term evaluated in the CM; these contributions have been
included in the results~\eqref{IijS}--\eqref{JijS} above.

Finally, we obtain for the flux (in the general orbit case but in the CM
frame) the structure
\begin{eqnarray}\label{fluxstruct}
\mathcal{F}&=&\frac{8}{15}\frac{G^3 m^4 \nu^2}{c^5
r^4}\left\{f_\mathrm{N}+\frac{1}{c^2}f_\mathrm{1PN}
+\frac{1}{c^3}\left[f_\mathrm{1.5PN}+\mathop{f}_{\text{S}}{}_{\!\mathrm{1.5PN}}
\right]+\frac{1}{c^4}\left[f_\mathrm{2PN}+\mathop{f}_{\text{SS}}{}_{\!\mathrm{2PN}}\right]
\right.\nonumber\\
&&\qquad\qquad\quad\left.+\frac{1}{c^5}\left[f_\mathrm{2.5PN}
+\mathop{f}_{\text{S}}{}_{\!\mathrm{2.5PN}} \right]+
\mathcal{O}\left(\frac{1}{c^6}\right)\right\}\,.
\end{eqnarray}
The non-spin pieces $f_\mathrm{N}$, $f_\mathrm{1PN}$, $f_\mathrm{1.5PN}$
and $f_\mathrm{2.5PN}$ are already known and we shall need below the
Newtonian and 1PN terms which are given by~\cite{WagW76,BS89}
\begin{subequations}\label{flux1PN}\begin{align}
f_\mathrm{N} &= 12v^2-11(nv)^2\,,\\f_\mathrm{1PN} &=
\left(\frac{785}{28}-\frac{213}{7}\nu\right)v^4
+\left(-\frac{1487}{14}+\frac{696}{7}\nu\right)(nv)^2v^2
+\left(\frac{2061}{28}-\frac{465}{7}\nu\right)(nv)^4\nonumber\\&+
\left(-\frac{680}{7}+\frac{40}{7}\nu\right)\frac{G\,m}{r}v^2
+\left(\frac{734}{7}-\frac{30}{7}\nu\right)\frac{G\,m}{r}(nv)^2
+\left(\frac{4}{7}-\frac{16}{7}\nu\right)\left(\frac{G\,m}{r}\right)^2\,.
\end{align}\end{subequations}
Notice also that the non-spin terms $f_\mathrm{1.5PN}$ and
$f_\mathrm{2.5PN}$ include the contributions of GW tails. Here we do not
deal with the spin-spin (SS) term at 2PN order which is given in
Refs.~\cite{KWW93,K95}. We obtain the SO coupling part at 1.5PN order as
\begin{eqnarray}\label{flux15PN}
\mathop{f}_{\text{S}}{}_{\!\mathrm{1.5PN}}&=&\frac{(S,n,v)}{m\,r}\left[78(nv)^2-8\frac{G
m}{r}-80
v^2\right]\nonumber\\&+&\frac{(\Sigma,n,v)}{m\,r}\left[51(nv)^2+4\frac{G
m}{r}-43v^2\right]\frac{\delta m}{m}\,,
\end{eqnarray}
and for this part we find perfect agreement with Kidder \textit{et al.}
\cite{KWW93,K95}. Finally, for the next-order SO part our result is
\begin{eqnarray}\label{flux25PN}
\mathop{f}_{\text{S}}{}_{\!\mathrm{2.5PN}}&=&\frac{(S,n,v)}{m\,r}\left[(nv)^4\left(
    - \frac{2244}{7}
 +\frac{3144}{7}\nu\right)+\frac{G^2m^2}{r^2}
  \left(\frac{972}{7}
+\frac{166}{7}\nu\right)\right.\nonumber\\ &&\qquad\qquad
  +\frac{G
    m}{r}(nv)^2\left(-\frac{2866}{7}+\frac{170}{7}\nu\right)+(nv)^2v^2\left(
  \frac{3519}{7}-\frac{5004}{7}\nu\right)\nonumber\\ &&\qquad\qquad
  \left.+\frac{G m}{r}v^2\left(\frac{3504}{7}-20\nu\right)
  +v^4\left(-\frac{1207}{7}+\frac{1810}{7}\nu\right)\right]\nonumber\\&+
&\frac{(\Sigma,n,v)}{m\,r}
\left[(nv)^4\left(-\frac{7941}{28} + \frac{2676}{7}\nu\right) +
\frac{G^2m^2}{r^2}\left(
  -\frac{109}{7}+18\nu\right)\right.\nonumber\\ &&\qquad\qquad +\frac{G
    m}{r}(nv)^2\left(-\frac{6613}{14}+\frac{1031}{7}\nu\right)+(nv)^2v^2\left(
  \frac{2364}{7} - \frac{3621}{7}\nu\right) \nonumber\\ &&\qquad\qquad \left. 
+\frac{G m}{r} v^2\left(\frac{4785}{14}-65\nu\right)
  +v^4\left(-\frac{2603}{28} + \frac{1040}{7}\nu\right)\right]\frac{\delta
  m}{m}\,.
\end{eqnarray}

\section{Reduction to quasi circular orbits}\label{secVI}

From now on we assume that when the binary enters the frequency
bandwidth of the LIGO/Virgo/LISA detectors the orbit has been
circularized by the gravitational radiation reaction effect. By
circular orbit we mean an orbit which is circular when the gradual
radiation reaction inspiral can be neglected, and when the effects of
spins are averaged over time. With such proviso there is a well
defined notion of a circular orbit (see Refs.~\cite{KWW93,K95} and
paper~I).

For circular orbits the orbital frequency $\omega$ is linked to the
distance $r$ between particles in harmonic coordinates by a relativistic
extension of Kepler's law, which has already been given in paper~I for
what concerns the SO effects. Let us write it again here, but let us
also add to it, for the benefit of potential users of these formulas,
all the non-spin contributions up to the 2.5PN order, following known
results from the literature (\textit{e.g.}~\cite{Bliving} and references
therein). The 3PN and 3.5PN non-spin terms, computed in
Refs.~\cite{BFIJ02,BDEI04}, can be added straightforwardly if necessary.
However, for convenience in this paper, we shall not display the
non-linear spin-spin (SS) terms. Thus, all formulas of this Section will
be complete up to 2.5PN order at \textit{linear} order in the spins
(\textit{i.e.} but for the SS contributions). We have
\begin{align}\label{om2}
\omega^2&=\frac{G\,m}{r^3}\,\left\{ 1
 +\gamma\left(-3+\nu\right)+\gamma^2\left(6+\frac{41}{4}\nu+\nu^2\right)
 \right.\nonumber\\&\qquad+\frac{\gamma^{3/2}}{G\,m^2}\left[-5S_\ell
 -3\frac{\delta
 m}{m}\Sigma_\ell\right]\nonumber\\&\left.\qquad+\frac{\gamma^{5/2}}{G\,m^2}\left[
 \left(\frac{39}{2}-\frac{23}{2}\nu\right)S_\ell
 +\left(\frac{21}{2}-\frac{11}{2}\nu\right)\frac{\delta
 m}{m}\Sigma_\ell\right]+ \mathcal{O}\left(\frac{1}{c^6}\right)\right\}\,,
\end{align}
in which $\gamma\equiv
\frac{G\,m}{r\,c^2}=\mathcal{O}\left(c^{-2}\right)$ denotes the
harmonic-coordinate PN parameter. We recognize the lowest-order (1.5PN
$\sim\gamma^{3/2}$) spin-orbit term and its 1PN correction at the
2.5PN $\sim\gamma^{5/2}$ level. Here, as in paper I, we introduce an
orthonormal triad $\{\mathbf{n},\bm{\lambda},\bm{\ell}\}$ defined by
$\mathbf{n}=\mathbf{x}/r$,
$\bm{\ell}=\mathbf{L}_\mathrm{N}/\vert\mathbf{L}_\mathrm{N}\vert$
where $\mathbf{L}_\mathrm{N}\equiv\mu\,\mathbf{x}\times\mathbf{v}$
denotes the Newtonian angular momentum, and
$\bm{\lambda}=\bm{\ell}\times\mathbf{n}$. The quantities $S_\ell$ and
$\Sigma_\ell$ in Eq.~(\ref{om2}) are the components of the spin
vectors~\eqref{SSigma} perpendicular to the orbital plane, namely
$S_\ell\equiv\mathbf{S}\cdot \bm{\ell}$ and $\Sigma_\ell\equiv
\mathbf{\Sigma}\cdot \bm{\ell}$. The relation~\eqref{om2} can be
inverted to give $\gamma$ in terms of an alternative PN parameter $x$,
directly related to the orbital frequency through
$x\equiv\left(\frac{G\,m\,\omega}{c^3}\right)^{2/3}=\mathcal{O}\left(c^{-2}\right)$.
As usual, it is better to express the PN formulas in terms of the
frequency-dependent PN parameter $x$ rather than $\gamma$ because they
are invariant under a large class of gauge transformations. Hence,
\begin{align}\label{gam}
\gamma&=x\,\left\{ 1
 +x\left(1-\frac{\nu}{3}\right)+x^2\left(1-\frac{65}{12}\nu\right)
 \right.\nonumber\\&\qquad+\frac{x^{3/2}}{G\,m^2}\left[\frac{5}{3}S_\ell
 +\frac{\delta
 m}{m}\Sigma_\ell\right]\nonumber\\&\left.\qquad+\frac{x^{5/2}}{G\,m^2}\left[
 \left(\frac{13}{3}+\frac{2}{9}\nu\right)S_\ell
 +\left(3-\frac{\nu}{3}\right)\frac{\delta m}{m}\Sigma_\ell\right]+
 \mathcal{O}\left(\frac{1}{c^6}\right)\right\}\,.
\end{align}
The SO term at order 1.5PN $\sim x^{3/2}$ is in agreement with Eq.~(16)
in~\cite{KWW93}.

The reduction of the GW flux $\mathcal{F}$, given by
Eqs.~\eqref{flux15PN}--\eqref{flux25PN}, to circular orbits is
straightforward, but care has to be taken from the fact that the
\textit{non-spin} parts of the flux at Newtonian and 1PN orders yield
crucial contributions to the SO terms for circular orbits [beside the
ones given by straightforward reduction of
Eqs.~\eqref{flux15PN}--\eqref{flux25PN}]. Such contributions are
generated by replacement of Eq.~\eqref{om2} into the 1PN flux given for
general orbits by Eq.~\eqref{flux1PN}. Finally, we obtain
\begin{align}\label{fluxgam}
\mathcal{F}=&\frac{32}{5}\frac{c^5}{G}\,\gamma^5\,\nu^2\left\{ 1
 +\gamma\left(-\frac{2927}{336}-\frac{5}{4}\nu\right)+4\pi\gamma^{3/2}
 \right.\nonumber\\&\qquad+\gamma^2\left(\frac{293383}{9072}
 +\frac{380}{9}\nu\right)
 +\pi\gamma^{5/2}\left(-\frac{25663}{672}-\frac{125}{8}\nu\right)
 \nonumber\\&\qquad+\frac{\gamma^{3/2}}{G\,m^2}\left[-\frac{37}{3}S_\ell
 -\frac{25}{4}\frac{\delta
 m}{m}\Sigma_\ell\right]\nonumber\\&\left.\qquad+\frac{\gamma^{5/2}}{G\,m^2}\left[
 \left(\frac{17897}{168}+23 \nu\right)S_\ell
 +\left(\frac{6253}{112} + \frac{277}{24}\nu\right)\frac{\delta
 m}{m}\Sigma_\ell\right]+
 \mathcal{O}\left(\frac{1}{c^6}\right)\right\}\,,
\end{align}
or, equivalently, in terms of the PN parameter $x$,
\begin{align}\label{fluxx}
\mathcal{F}=&\frac{32}{5}\frac{c^5}{G}\,x^5\,\nu^2\left\{ 1
 +x\left(-\frac{1247}{336}-\frac{35}{12}\nu\right)+4\pi x^{3/2}
 \right.\nonumber\\&\qquad+x^2\left(-\frac{44711}{9072}+\frac{9271}{504}\nu
 +\frac{65}{18}\nu^2\right) +\pi
 x^{5/2}\left(-\frac{8191}{672}-\frac{583}{24}\nu\right)
 \nonumber\\&\qquad+\frac{x^{3/2}}{G\,m^2}\left[-4S_\ell
 -\frac{5}{4}\frac{\delta
 m}{m}\Sigma_\ell\right]\nonumber\\&\left.\qquad+\frac{x^{5/2}}{G\,m^2}\left[
 \left(-\frac{23}{4}+\frac{245}{9}\nu\right)S_\ell
 +\left(-\frac{33}{16} + \frac{37}{4}\nu\right)\frac{\delta
 m}{m}\Sigma_\ell\right]+
 \mathcal{O}\left(\frac{1}{c^6}\right)\right\}\,.
\end{align}
All the non-spin terms are included up to 2.5PN order. Notice in
particular the non-spin terms, proportional to $\pi$, which are at the
same 1.5PN and 2.5PN orders as the SO effects; these terms are due to GW
tails~\cite{BFIJ02,Bliving}.\,\footnote{For the non-spin tail term at
2.5PN order we take into account the published Erratum
to~\cite{BFIJ02}.} For the leading SO term at order 1.5PN $\sim x^{3/2}$
we find perfect agreement with Eq.~(17b) in~\cite{KWW93}. We also check that
the result agrees in the test-mass limit $\nu \to 0$ with the black-hole
perturbation calculation of Tagoshi \emph{et al.} \cite{TSTS96} [see Eq.~(G19)
there].

The reduction of the center-of-mass energy $E$ (computed in Sec.~VII of
paper~I) to circular orbits is straightforward, and we simply report
here the final result, completing it by the known non-spin terms \cite{BI03CM}.
We have
\begin{align}\label{Egam}
E=&-\frac{\mu\,c^2\,\gamma}{2}\,\left\{ 1
 +\gamma\left(-\frac{7}{4}+\frac{\nu}{4}\right)+\gamma^2\left(-\frac{7}{8}
 +\frac{49}{8}\nu+\frac{\nu^2}{8}\right)
 \right.\nonumber\\&\qquad+\frac{\gamma^{3/2}}{G\,m^2}\left[ 3S_\ell
 +\frac{\delta
 m}{m}\Sigma_\ell\right]\nonumber\\&\left.\qquad+\frac{\gamma^{5/2}}{G\,m^2}\left[
 \left(7-4\nu\right)S_\ell +\left(3 -2\nu\right)\frac{\delta
 m}{m}\Sigma_\ell\right]+
 \mathcal{O}\left(\frac{1}{c^6}\right)\right\}\,,
\end{align}
or, equivalently,
\begin{align}\label{Ex}
E=&-\frac{\mu\,c^2\,x}{2}\,\left\{ 1
 +x\left(-\frac{3}{4}-\frac{\nu}{12}\right)+x^2\left(-\frac{27}{8}
 +\frac{19}{8}\nu-\frac{\nu^2}{24}\right)
 \right.\nonumber\\&\qquad+\frac{x^{3/2}}{G\,m^2}\left[
 \frac{14}{3}S_\ell +2\frac{\delta
 m}{m}\Sigma_\ell\right]\nonumber\\&\left.\qquad+\frac{x^{5/2}}{G\,m^2}\left[
 \left(13-\frac{49}{9}\nu\right)S_\ell
 +\left(5-\frac{8}{3}\nu\right)\frac{\delta m}{m}\Sigma_\ell\right]+
 \mathcal{O}\left(\frac{1}{c^6}\right)\right\}\,.
\end{align}
Alternatively, in terms of the single-spin variables the spin-dependent
part of the above equation reads
\begin{align}\label{ExS}
\mathop{E}_\text{S} &=-\frac{\mu\,c^2\,x}{2}\,\sum_{i = 1,2}\chi_i\,
\kappa_i\left\{ x^{3/2} \left[\frac{8}{3}\frac{m_i^2}{m^2} + 2\nu\right]
\right.\nonumber \\ & \qquad\qquad\left. + {x^{5/2}} \left[
\frac{m_i^2}{m^2}\,\left( 8 - \frac{25}{9}\nu \right) + \nu \left(5-
\frac{8}{3}\nu \right) \right] \right\}\,,
\end{align}
where we denote by $\kappa_i= \hat{\mathbf{S}}_i \cdot \bm{\ell}$ for
$i=1,2$ the orientation of the spins with respect to the Newtonian
angular momentum, and by $\chi_i$ their magnitude defined in the
standard way by $\mathbf{S}_i=G \,m_i^2\,\chi_i\,\hat{\mathbf{S}}_i$.

Assuming non-precessing orbits, we list in Table~\ref{Tableico} the
energy and the frequency at the so-called
innermost circular orbit (ICO)~\cite{B01}. The ICO is defined by the
minimum of the center-of-mass energy for circular orbits expressed as a
function of the orbital frequency $\omega$, and is computed from
Eq.~\eqref{Ex}.
\begin{table}[h]
\begin{tabular}{ll||c||c}
&& $m\,\omega_\mathrm{ICO}$ & $E_\mathrm{ICO}/m$ \\ \hline 1PN &
&$0.522$& $-0.0405$ \\\hline &$\kappa_i=0$ & $0.522$ & $-0.0405$ \\1.5PN
&$\kappa_i=+1$ & $-$ & $-$ \\ &$\kappa_i=-1$ & $0.111$ & $-0.0163$
\\\hline &$\kappa_i=0$& $0.137$ & $-0.0199$ \\ 2PN
&$\kappa_i=+1$&$0.318$ & $-0.0390$ \\ &$\kappa_i=-1$& $0.0733$ &$
-0.0130$ \\ \hline &$\kappa_i=0$& $0.137$ & $-0.0199$ \\2.5PN
&$\kappa_i=+1$ & $-$ & $-$ \\ &$\kappa_i=-1$ & $0.060$ & $-0.0117$
\\\hline &$\kappa_i=0$& $0.129$ & $-0.0193$ \\3PN &$\kappa_i=+1$ & $-$ &
$-$ \\ &$\kappa_i=-1$ & $0.059$ & $-0.0116$ \\\hline
\end{tabular}
\caption{Energy and angular frequency at the
ICO for equal-mass ($\nu=\frac{1}{4}$) binary systems. The spins are
maximal ($\chi_i=1$) and have different orientations ($\kappa_i=0,\,\pm
1$). In three cases, indicated by a dash, there is no ICO, \textit{i.e.}
the energy function admits no real minimum. Spin-spin effects at 2PN
order are included.}
\label{Tableico}
\end{table}

The orbital angular momentum (computed in paper~I) in the case of
circular orbits reads
\begin{align}
\mathbf{L} &= \frac{G\,m^2}{c}\,\nu\, \gamma^{-1/2}\,\bigg\{\bm{\ell}
 \bigg [ (1+2\gamma) + \bigg( -3 \,S_\ell -
\Sigma_\ell \, \frac{\delta m}{m} \bigg) \,\frac{\gamma^{3/2}}{G\,m^2}
\nonumber \\ & \qquad +
\bigg(\frac{5}{2} - \frac{9}{2} \nu \bigg) \,\gamma^2 +
 \bigg( \Big( - \frac{59}{8} +
\frac{25}{8}\nu \Big)  \,S_\ell +  \Big( - \frac{27}{8} + \frac{3}{2}\nu
\Big) \,\frac{\delta 
m}{m}\, \Sigma_\ell \bigg) \frac{\gamma^{5/2}}{G\,m^2}
\bigg] \nonumber \\ & + \frac{\gamma^{3/2}}{G m^2} \, \boldsymbol{\lambda}
\bigg[  - \frac{3}{2} \, S_\lambda - \frac{1}{2} \, \Sigma_\lambda \,
\frac{\delta m}{m} + \bigg( \Big(\frac{3}{8} - \frac{61}{8} \nu \Big) \,
S_\lambda + \Big(-\frac{9}{8} - \frac{15}{4} \nu \Big)  \,
\frac{\delta m}{m} \, \Sigma_\lambda \bigg) \gamma \bigg] \nonumber \\ &
+ \frac{\gamma^{3/2}}{G m^2} \, \mathbf{n} \bigg[ \frac{5}{2} \, S_n  +
\frac{3}{2} \, \Sigma_n  \, \frac{\delta m}{m} + \bigg(  \Big(- \frac{13}{8} -
\frac{13}{8} \nu \Big) \, S_n +  \Big( \frac{15}{8} - \frac{3}{4} \nu
\Big)   \, \frac{\delta m}{m} \,\Sigma_n \bigg) \gamma \bigg] 
\bigg\} \,, 
\end{align}
or equivalently
\begin{align}
\mathbf{L} &= \frac{G\,m^2}{c}\,\nu\,\,x^{-1/2}\,\bigg\{ \bm{\ell}
 \bigg[ 1 + \bigg( \frac{3}{2} + \frac{\nu}{6} \bigg) \,x +
\bigg( - \frac{23}{6}S_\ell - \frac{3}{2}\, \frac{\delta m}{m}\,\Sigma_\ell
\bigg)  \,\frac{x^{3/2}}{G\,m^2}
\nonumber \\ &  \qquad + \bigg( \frac{27}{8} - \frac{19
\nu}{8} + \frac{\nu^2}{24} \bigg)\,x^2 + \bigg(  \Big( -
\frac{77}{8} + \frac{259}{72}\nu \Big) \,S_\ell +\Big( - \frac{33}{8} +
\frac{7}{4}\nu \Big) \,\frac{\delta m}{m}\,\Sigma_\ell \bigg)
\,\frac{x^{5/2}}{G\,m^2} \nonumber \\ & +
\frac{x^{3/2}}{G m^2} \, \boldsymbol{\lambda} \bigg[ - \frac{3}{2}\, S_\lambda
- \frac{1}{2}  \, \Sigma_\lambda \, \frac{\delta m}{m} + \bigg( \Big( -
\frac{9}{8} - \frac{57}{8} \nu \Big) \,
S_\lambda + \Big(- \frac{13}{8} - \frac{43}{12} \nu \Big)  \,\frac{\delta
  m}{m}\, \Sigma_\lambda \bigg) x
\bigg] \nonumber \\ & + \frac{x^{3/2}}{G m^2} \, \mathbf{n}
\bigg[ \frac{5}{2}\, S_n
+ \frac{3}{2} \, \Sigma_n \, \frac{\delta m}{m} + \bigg( \Big( 
\frac{7}{8} - \frac{59}{24} \nu \Big) \,
S_n + \Big(\frac{27}{8} - \frac{5}{4} \nu \Big)  \,\frac{\delta m}{m}\,
\Sigma_n \bigg) x 
\bigg] \bigg\}\,. 
\end{align}

For future use we give here the precessional equations evaluated in
paper~I, but reduced to circular orbits:
\begin{subequations}\label{dSSigma}\begin{align}
\frac{d \mathbf{S}}{d t} &=
\nu\,\omega\left\{x\left[-4S_\lambda-2\frac{\delta
m}{m}\,\Sigma_\lambda\right]\mathbf{n}+x\left[3S_n+\frac{\delta
m}{m}\,\Sigma_n\right]\bm{\lambda}\right.\nonumber\\&\qquad\quad\left.
+x^2\left[\left(1
-\frac{20}{3}\nu\right)S_\lambda+\left(-2-\frac{10}{3}\nu\right)\frac{\delta
m}{m}\,\Sigma_\lambda\right]\mathbf{n}\right.
\nonumber\\&\qquad\quad\left.+x^2\left[\left(9-11\nu\right)S_n
-\frac{16}{3}\nu\,\frac{\delta
m}{m}\,\Sigma_n\right]\bm{\lambda}\right\}+
\mathcal{O}\left(\frac{1}{c^6}\right)\,,\\ \frac{d \mathbf{\Sigma}}{d t}
&= \omega\left\{x\left[\left(-2+4\nu\right)\Sigma_\lambda-2\frac{\delta
m}{m}\,S_\lambda\right]\mathbf{n}+x\left[\left(1-\nu\right)\Sigma_n+\frac{\delta
m}{m}\,S_n\right]\bm{\lambda}\right.\nonumber\\&\qquad\quad\left.
+x^2\left[\left(-2
+\frac{17}{3}\nu+\frac{20}{3}\nu^2\right)\Sigma_\lambda+\left(-2
-\frac{10}{3}\nu\right)\frac{\delta
m}{m}\,S_\lambda\right]\mathbf{n}\right.
\nonumber\\&\qquad\quad\left.+x^2\left[\left(\frac{11}{3}\nu
+\frac{31}{3}\nu^2\right) \Sigma_n-\frac{16}{3}\nu\,\frac{\delta
m}{m}\,S_n\right]\bm{\lambda}\right\}+
\mathcal{O}\left(\frac{1}{c^6}\right)\,.
\end{align}\end{subequations}
We recall our notation~\eqref{SSigma} for the spin variables. We denote
by $S_n$, $\Sigma_n$ and $S_\lambda$, $\Sigma_\lambda$ the components of
the spins along the vectors $\mathbf{n}$ and $\bm{\lambda}$
respectively. As before we have neglected the SS terms.

To compare easily with previous results in the literature, we have also
computed the precessing equations for the spin variables $\mathbf{S}_1$
and $\mathbf{S}_2$. They read
\begin{subequations}\label{our}\begin{align}
\label{our1}
\frac{d \mathbf{S}_1}{d t} &= \omega\,\nu\,x\,\left \{
S_{1n}\,\bm{\lambda}\,\left (2+ \frac{m_2}{m_1}\right ) -2
S_{1\lambda}\,\mathbf{n}\,\left (1+ \frac{m_2}{m_1}\right ) \right.
\nonumber \\ & \left. + x\,S_{1n}\,\bm{\lambda}\,\left
(9\,\frac{m_1^2}{m^2} + \frac{37}{3}\, \frac{m_1\,m_2}{m^2} +
\frac{11}{3}\,\frac{m_2^2}{m^2} \right ) \right . \nonumber \\ & \left.
+ x\,S_{1\lambda}\,\mathbf{n}\,\left (3\,\frac{m_1}{m} -
\frac{7}{3}\,\frac{m_2}{m} - 2\,\frac{m_2^2}{m\,m_1} \right ) \right \}+
\mathcal{O}\left(\frac{1}{c^6}\right)\,, \\
\label{our2}
\frac{d \mathbf{S}_2}{d t} &= \omega\,\nu\,x\,\left \{
S_{2n}\,\bm{\lambda}\,\left (2+ \frac{m_1}{m_2}\right ) -2
S_{2\lambda}\,\mathbf{n}\,\left (1+ \frac{m_1}{m_2}\right ) \right.
\nonumber \\ & \left . + x\,S_{2n}\,\bm{\lambda}\,\left
(9\,\frac{m_2^2}{m^2} + \frac{37}{3}\, \frac{m_1\,m_2}{m^2} +
\frac{11}{3}\,\frac{m_1^2}{m^2} \right ) \right . \nonumber \\ & \left
. + x\,S_{2\lambda}\,\mathbf{n}\,\left (3\,\frac{m_2}{m} -
\frac{7}{3}\,\frac{m_1}{m} - 2\,\frac{m_1^2}{m\,m_2} \right ) \right \}+
\mathcal{O}\left(\frac{1}{c^6}\right)\,,
\end{align}\end{subequations}
where $S_{1\lambda}, S_{1n}$ and $S_{2\lambda}, S_{2n}$ are the
projections of the single-spin variables along $\bm{\lambda}$ and
$\mathbf{n}$. We notice that with our choice of spin variables
$\mathbf{S}_1$ and $\mathbf{S}_2$ the magnitude of the spin is not
constant even when restricting Eqs.~\eqref{our} to 1.5PN order. In
Sec.~\ref{secVII} we define some alternative spin variables
$\mathbf{S}^\text{c}_1$ and $\mathbf{S}^\text{c}_2$, such that the
magnitude of these spin vectors remains constant, \textit{i.e.}
$\mathbf{S}^\text{c}_i \cdot {d \mathbf{S}^\text{c}_i}/{d t}=0$ with
$i=1,2$. The spin vectors $\mathbf{S}^\text{c}_i$ agree with
Kidder's~\cite{K95} spin variables at the 1PN order, and generalize them
to the next 2PN order. The main advantage of the definition
$\mathbf{S}^\text{c}_i$ is that the precession equations can be then
written in the form
$d\mathbf{S}^\text{c}_i/dt=\mathbf{\Omega}_i\times\mathbf{S}^\text{c}_i$,
where $\mathbf{\Omega}_i$ are the precession angular frequency vectors
(given in Sec.~\ref{secVII}).

\section{Spin variables with constant magnitude}\label{secVII}

In this paper and paper~I we found convenient to use some specific spin
variables $\mathbf{S}_1$ and $\mathbf{S}_2$, defined in Sec.~II of paper
I. However, as discussed in paper~I, other papers in the literature use
a definition of the spin variables different from ours. 
For example, the spin-precession equations at 1PN order in Ref.~\cite{K95} read
\begin{subequations}\label{kidderdS}\begin{eqnarray}
\label{kidderdS1}
\frac{d \mathbf{S}^\text{c}_1}{d t} &=& \omega\,\nu\,x\,(\bm{\ell}
\times \mathbf{S}^\text{c}_1)\, \left ( 2 + \frac{3}{2}\frac{m_2}{m_1}
\right )= \omega\,\nu\,x \left [ S^\text{c}_{1n}\,\bm{\lambda} -
S^\text{c}_{1\lambda}\,\mathbf{n}\right ] \,\left ( 2 +
\frac{3}{2}\frac{m_2}{m_1} \right )\,, \\
\label{kidderdS2}
\frac{d \mathbf{S}^\text{c}_2}{d t} &=& \omega\,\nu\,x\,(\bm{\ell}
\times \mathbf{S}^\text{c}_2)\, \left ( 2 + \frac{3}{2}\frac{m_1}{m_2}
\right ) = \omega\,\nu\,x \left [ S^\text{c}_{2n}\,\bm{\lambda} -
S^\text{c}_{2\lambda}\,\mathbf{n}\right ] \,\left ( 2 +
\frac{3}{2}\frac{m_1}{m_2} \right )\,,
\end{eqnarray}\end{subequations}
where the superscript c stands for constant; in fact, the spin variables
$\mathbf{S}^\text{c}_1$, $\mathbf{S}^\text{c}_2$ are such that their
norm or magnitude remains constant. Indeed one can readily check from
Eqs.~\eqref{kidderdS} that $\mathbf{S}^\text{c}_i \cdot {d
\mathbf{S}^\text{c}_i}/{d t}=0$ with $i=1,2$. Our spin variables are
related at the 1PN order to the constant-spin ones (in the
center-of-mass and for circular orbits) as
\begin{subequations}\label{kidderS}\begin{align}\label{kidderS1}
\mathbf{S}^\text{c}_1 &=\left(1+\frac{G
m_2}{c^2r}\right)\mathbf{S}_1-\frac{m_2^2}{2c^2 m^2}\,S_{1\lambda}
\,r^2\,\omega^2\,\bm{\lambda}\,, \\
\label{kidderS2}
\mathbf{S}^\text{c}_2 &=\left(1+\frac{Gm_1}{c^2r}\right)\mathbf{S}_2-
\frac{m_1^2}{2c^2 m^2}\,S_{2\lambda}\,r^2\,\omega^2\,\bm{\lambda}\,.
\end{align}\end{subequations}
We can check that by taking the time derivative of the RHS of
Eqs.~\eqref{kidderS}, plugging in Eqs.~\eqref{our} at 1PN order, we
recover Eqs.~\eqref{kidderdS}. Note that the total angular momentum is
invariant, since
\begin{equation}
\mathbf{J} = \mathbf{L} + \frac{1}{c} \mathbf{S}_1 + \frac{1}{c} \mathbf{S}_2
= \mathbf{L}^\text{c} + \frac{1}{c} \mathbf{S}^\text{c}_1 + \frac{1}{c}
\mathbf{S}^\text{c}_2 \, .
\end{equation}%

Let us now define, at the 2PN order, in a general frame and for
general orbits, some spin variables reducing to
$\mathbf{S}^\text{c}_1$ and $\mathbf{S}^\text{c}_2$ at the 1PN order,
and such that the magnitude of these spins remains constant. We shall
still denote the latter 2PN spins as $\mathbf{S}^\text{c}_1$,
$\mathbf{S}^\text{c}_2$; thus, we shall have, at the 2PN order,
$\mathbf{S}^\text{c}_i \cdot {d \mathbf{S}^\text{c}_i}/{d t}=0$ with
$i=1,2$. First of all we find that the new spin variables are related
to the ones used in previous sections (and in the whole of paper~I) by
\begin{align}\label{S1K}
\mathbf{S}_1^\text{c} &= \mathbf{S}_1 + \frac{1}{c^2} \bigg[ -
\frac{1}{2} (v_1 S_1) \mathbf{v}_1 + \frac{G m_2}{r_{12}} \mathbf{S}_1
\bigg] \nonumber \\ & + \frac{1}{c^4} \bigg[\mathbf{n}_{12} \frac{G
m_2}{r_{12}} (n_{12} S_1) \bigg(-4 \frac{G m_1}{r_{12}} + \frac{1}{2}
\frac{G m_2}{r_{12}} \bigg) + \frac{1}{2} \frac{G m_2}{r_{12}}
\mathbf{S}_1 \Big(- (n_{12}v_2)^2 + \frac{G m_1}{r_{12}} \Big) \nonumber
\\ & \qquad + \mathbf{v}_1 \bigg(- \frac{1}{8} (v_1S_1) v_1^2 + \frac{G
m_2}{r_{12}} \Big( -\frac{5}{2} (v_1S_1) + 4 (v_2S_1) \Big) \bigg) + 2
\frac{G m_2}{r_{12}} \mathbf{v}_2 (v_2S_1)\bigg]\,,
\end{align}
together with the expression for $\mathbf{S}_2^\text{c}$ obtained by
exchanging all the particle's labels $1\leftrightarrow 2$. In
Eq.~\eqref{S1K} the notation is exactly the same as in paper~I. The main
advantage of such definition~\eqref{S1K} is that the precession
equations can now be written into the form
\begin{subequations}\label{preceq}\begin{align}
\frac{d \mathbf{S}_1^\text{c}}{d t} &= \mathbf{\Omega}_1\times
\mathbf{S}_1^\text{c}\,,\\ \frac{d \mathbf{S}_2^\text{c}}{d t} &=
\mathbf{\Omega}_2\times \mathbf{S}_2^\text{c}\,,
\end{align}\end{subequations}
showing that the spins precess around the directions of
$\mathbf{\Omega}_1$ and $\mathbf{\Omega}_2$, and at the rates
$\vert\mathbf{\Omega}_1\vert$ and $\vert\mathbf{\Omega}_2\vert$. The
precession angular frequency vectors $\mathbf{\Omega}_1$,
$\mathbf{\Omega}_2$ can be computed up to the 2PN order by using the
precession equations~(6.1)--(6.3) of paper~I. We find
\begin{align}\label{Omega1}
\mathbf{\Omega}_1 &= \frac{G m_2}{c^2 r_{12}^2} \bigg[ \frac{3}{2}
\mathbf{n}_{12}\times \mathbf{v}_1 - 2 \mathbf{n}_{12}\times
\mathbf{v}_2\bigg] \nonumber \\ & + \frac{G m_2}{c^4 r_{12}^2} \bigg[
\mathbf{n}_{12}\times \mathbf{v}_1 \Big(-\frac{9}{4} (n_{12}v_2)^2 +
\frac{1}{8} v_1^2 - (v_1v_2) + v_2^2 + \frac{7}{2} \frac{G m_1}{r_{12}}
- \frac{1}{2} \frac{G m_2}{r_{12}} \Big) \nonumber \\ & \qquad\quad +
\mathbf{n}_{12}\times \mathbf{v}_2 \Big(3 (n_{12}v_2)^2 + 2 (v_1v_2) - 2
v_2^2 + \frac{G m_1}{r_{12}} + \frac{9}{2} \frac{G m_2}{r_{12}} \Big)
\nonumber \\ & \qquad\quad + \mathbf{v}_1 \times \mathbf{v}_2 \Big( 3
(n_{12}v_1) - \frac{7}{2} (n_{12}v_2) \Big)\bigg]\,,
\end{align}
together with $1\leftrightarrow 2$. In the center-of-mass frame we
obtain
\begin{align} \label{Omega1cm}
\mathbf{\Omega}_1 &= \frac{G m}{r^2 c^2} \biggl\{ \frac{3}{4} +
  \frac{\nu}{2} + \frac{1}{c^2} \bigg[\frac{G m}{r} \Big( -\frac{1}{4} -
  \frac{3}{8} \nu + \frac{\nu^2}{2} \Big) \nonumber \\ & \qquad \qquad +
  \Big( - \frac{3}{2} \nu + \frac{3}{4} \nu^2 \Big) (nv)^2 + \Big(
  \frac{1}{16} + \frac{11}{8} \nu - \frac{3}{8} \nu^2 \Big) v^2 \bigg]
  \nonumber \\ & \qquad + \frac{\delta m}{m}\left(-\frac{3}{4} +
  \frac{1}{c^2} \bigg[ \frac{G m}{r} \Big(\frac{1}{4} - \frac{\nu}{8}
  \Big) - \frac{3}{2} \nu (nv)^2 + \Big(-\frac{1}{16} +
  \frac{\nu}{2}\Big) v^2 \bigg]\right) \biggr\}\,\mathbf{n}\times
  \mathbf{v}\,.
\end{align}
For circular orbits,
\begin{equation} \label{Omega1circ}
\mathbf{\Omega}_1 = \frac{c^3 x^{5/2}}{G m} \biggl\{ \frac{3}{4} +
  \frac{\nu}{2} + x \Big( \frac{9}{16} + \frac{5}{4} \nu -
  \frac{\nu^2}{24} \Big) + \frac{\delta m}{m}\left[ -\frac{3}{4} + x
  \Big( -\frac{9}{16} + \frac{5}{8}\nu
  \Big)\right]\biggr\}\,\bm{\ell}\,.
\end{equation}
Note that to obtain $\mathbf{\Omega}_2$ we simply have to change $\delta
m\rightarrow -\delta m$. Recall that all these expressions are valid at
linear order in the spins (excluding the SS terms); this means that with
this approximation $\mathbf{\Omega}_1$ and $\mathbf{\Omega}_2$ are
independent of the spins.

Finally let us express some of the main results of this paper in terms
of the spin variables with constant magnitude. For the spin-dependent
part of the circular-orbit energy [Eqs.~\eqref{Ex} or~\eqref{ExS}] we
get
\begin{align}\label{ExK}
\mathop{E}_{\text{S}} =&-\frac{\mu\,c^2\,x}{2}\,\left\{ 1
+\frac{x^{3/2}}{G\,m^2}\left[ \frac{14}{3}S^\text{c}_\ell +2\frac{\delta
m}{m}\Sigma^\text{c}_\ell\right] \right. \nonumber\\
&\left.\qquad+\frac{x^{5/2}}{G\,m^2}\left[
\left(11-\frac{61}{9}\nu\right)S^\text{c}_\ell
+\left(3-\frac{10}{3}\nu\right)\frac{\delta
m}{m}\Sigma^\text{c}_\ell\right]+
\mathcal{O}\left(\frac{1}{c^6}\right)\right\}\,,
\end{align}
where $S^\text{c}_\ell\equiv\mathbf{S}^\text{c}\cdot \bm{\ell}$ and
$\Sigma^\text{c}_\ell\equiv \mathbf{\Sigma}^\text{c}\cdot \bm{\ell}$,
with $\mathbf{S}^\text{c}=\mathbf{S}^\text{c}_1+\mathbf{S}^\text{c}_2$
and $\mathbf{\Sigma}^\text{c}=\frac{m}{m_2}\mathbf{S}^\text{c}_2
-\frac{m}{m_1}\mathbf{S}^\text{c}_1$. Using the energy~\eqref{ExK} as
function of the constant spin variables, we have computed the ICO. For
an equal-mass binary with spins anti-aligned with the orbital angular
momentum, at 2.5PN (3PN) order, we get $E_\mathrm{ICO}/m = -0.0122$
($E_\mathrm{ICO}/m = -0.0119$) and $m \omega_\mathrm{ICO} = 0.064$ ($m
\omega_\mathrm{ICO} = 0.061$) to be compared with the numbers listed
in Table~\ref{Tableico}. The difference is not negligible.

We also computed the spin-dependent part of the orbital angular
momentum and the flux in terms of the spin variables with constant
magnitude,
\begin{align}\label{LxK}
\mathop{\mathbf{L}^\text{c}}_\text{S} &=
 \frac{G\,m^2}{c}\,\nu\,\,x^{-1/2}\,\bigg\{ \bm{\ell} \bigg[ \bigg( -
 \frac{35}{6}S^\text{c}_\ell - \frac{5}{2}\, \frac{\delta
 m}{m}\,\Sigma^\text{c}_\ell \bigg) \,\frac{x^{3/2}}{G\,m^2} \nonumber
 \\ & \qquad + \bigg( \Big( - \frac{77}{8} + \frac{427}{72}\nu \Big)
 \,S^\text{c}_\ell +\Big( - \frac{21}{8} + \frac{35}{12}\nu \Big)
 \,\frac{\delta m}{m}\,\Sigma^\text{c}_\ell \bigg)
 \,\frac{x^{5/2}}{G\,m^2} \bigg] \nonumber \\ & + \frac{x^{3/2}}{G
 m^2} \, \boldsymbol{\lambda} \bigg[ -{3}\, S^\text{c}_\lambda - \,
 \Sigma^\text{c}_\lambda \, \frac{\delta m}{m} + \bigg( \Big( -
 \frac{7}{2} +3 \nu \Big) \, S^\text{c}_\lambda + \Big(- \frac{1}{2} +
 \frac{4}{3} \nu \Big) \,\frac{\delta m}{m}\, \Sigma^\text{c}_\lambda
 \bigg) x \bigg] \nonumber \\ & + \frac{x^{3/2}}{G m^2} \, \mathbf{n}
 \bigg[ \frac{1}{2}\, S^\text{c}_n + \frac{1}{2} \, \Sigma^\text{c}_n
 \, \frac{\delta m}{m} + \bigg( \Big( \frac{11}{8} - \frac{19}{24} \nu
 \Big) \, S^\text{c}_n + \Big(\frac{11}{8} - \frac{5}{12} \nu \Big)
 \,\frac{\delta m}{m}\, \Sigma^\text{c}_n \bigg)x \bigg] \bigg \}\,,
 \\
\label{FxK}\mathop{\mathcal{F}}_{\text{S}} &=\frac{32}{5}
\frac{c^5}{G}\,x^5\,\nu^2\left\{ 1
+\frac{x^{3/2}}{G\,m^2}\left[-4S^\text{c}_\ell -\frac{5}{4}\frac{\delta
m}{m}\Sigma^\text{c}_\ell\right] \right.
\nonumber\\&\left.\qquad+\frac{x^{5/2}}{G\,m^2}\left[
\left(-\frac{9}{2}+\frac{272}{9}\nu\right)S^\text{c}_\ell
+\left(-\frac{13}{16}+\frac{43}{4}\nu\right)\frac{\delta
m}{m}\Sigma^\text{c}_\ell\right]+
\mathcal{O}\left(\frac{1}{c^6}\right)\right\}\,.
\end{align}

\section{Phase evolution and accumulated number of GW cycles}\label{secVIII}

In this Section we compute the time evolution of the binary's orbital
frequency $\omega$, which results from the gravitational radiation
reaction damping force. Instead of computing directly the radiation
reaction force, we use the standard energy balance argument
\begin{equation}\label{balance}
\mathcal{F}=-\frac{dE}{dt}\,,
\end{equation}
where $\mathcal{F}$ is the total emitted GW energy flux computed in
Sec.~\ref{secV}, and $E$ denotes the binary's center-of-mass energy,
namely the integral of the motion associated with the conservative
part of the equations of motion, and which has been computed in
paper~I. Using $\mathcal{F}[x]$ and $E[x]$ expressed in terms of the
spin variables with constant magnitude by Eqs.~(\ref{ExK})
and~(\ref{FxK}), we deduce $\dot{\omega}$ from~\eqref{balance}, which
is equivalent to
\begin{equation}\label{balancex}
\frac{\dot{\omega}}{\omega}= -\frac{3}{2x}\,\left(\frac{d
  E\left[x\right]}{d x}\right)^{\!-1}\!\mathcal{F}\left[x\right]\,.
\end{equation}
Notice that for this calculation it is crucial to use the variables
associated with the constant magnitude spins, $\mathbf{S}^\text{c}$
and $\mathbf{\Sigma}^\text{c}$ (rather than our original spin
variables $\mathbf{S}$ and $\mathbf{\Sigma}$
\footnote{We are grateful to A. Gopakumar and G. Sch{\"a}fer for
pointing out this fact to us.}), since the spin variables
$\mathbf{S}^\text{c}$ and $\mathbf{\Sigma}^\text{c}$ are secularly
constant, \textit{i.e.} do not evolve by gravitational radiation
reaction. A proof that $\mathbf{S}^\text{c}$ and
$\mathbf{\Sigma}^\text{c}$ are secularly constant (to the considered
order) can be found in Ref.~\cite{W05}. Hence the balance equation in
the form of Eq.~(\ref{balancex}) gives directly the secular evolution
of the orbital frequency.

During this computation the standard PN approximation is applied,
\textit{i.e.} we expand both the numerator and the denominator of
Eq.~\eqref{balancex} in the usual PN way, and finally express the result
as a Taylor series in $x$. Other ways of addressing the computation,
using particular PN resummation techniques, can be found in
Refs.~\cite{BCV03a,BCV03b} and references therein. We give the end
result for the parameter $\xi\equiv\dot{\omega}/\omega^2$, which can be
viewed as the dimensionless adiabatic parameter associated with the
gradual inspiral, and which is dominantly of 2.5PN order (namely,
the order of radiation reaction). In the final result, as everywhere
else, the SO effects are at the 1.5PN and 2.5PN orders beyond the
dominant approximation. We get
\begin{align}\label{omdot}
\frac{\dot{\omega}}{\omega^2}=&\frac{96}{5}\,\nu\,x^{5/2}\left\{ 1
 +x\left(-\frac{743}{336}-\frac{11}{4}\nu\right)+4\pi x^{3/2}
 \right.\nonumber\\&\qquad+x^2\left(\frac{34103}{18144}+\frac{13661}{2016}\nu
 +\frac{59}{18}\nu^2\right) +\pi
 x^{5/2}\left(-\frac{4159}{672}-\frac{189}{8}\nu\right)
 \nonumber\\&\qquad+\frac{x^{3/2}}{G\,m^2}\left[-\frac{47}{3}S^\text{c}_\ell
 -\frac{25}{4}\frac{\delta
 m}{m}\Sigma^\text{c}_\ell\right]\nonumber\\&\left.\qquad
 +\frac{x^{5/2}}{G\,m^2}\left[
 \left(-\frac{5861}{144}+\frac{1001}{12}\nu\right)S^\text{c}_\ell
 +\left(-\frac{809}{84}+\frac{281}{8}\nu\right)\frac{\delta
 m}{m}\Sigma^\text{c}_\ell\right]\right\}\,.
\end{align}
If necessary the non-spin contributions at orders 3PN and 3.5PN
can be straightforwardly added~\cite{BFIJ02,BDEI04}.
 
Equations (\ref{preceq}), (\ref{Omega1circ}), (\ref{ExK}), (\ref{FxK})
and (\ref{omdot}) with non-spin terms added through 3.5PN order and
spin-spin terms included, together with the equation describing the
rate of change of the orbital angular-momentum direction (deduced from
$\dot{\mathbf{L}}^\text{c}=-\frac{1}{c}\dot{\mathbf{S}}_1^\text{c}
-\frac{1}{c}\dot{\mathbf{S}}_2^\text{c}$ at leading order)
and the radiation field (see, e.g., Sec.~C and Appendix~B in
Ref.~\cite{K95}), can be solved semi-analytically, for special spin and
mass configurations, or numerically. They provide more accurate
templates than currently used in the literature
~\cite{3mn,ACST94,A95,A96a,A96b,GKV03,BCV03b,PBCV04,
GK03,BCPV04,GIKB04,BCPTV05} and would need to be implemented for the
search of GWs from spinning, precessing binaries.

In the general case, taking into account the effect of precession of the
orbital plane induced by spin modulations, the GW phase
$\Phi_\mathrm{GW}$ is given by $\Phi_\mathrm{GW} = \phi_\mathrm{GW} +
\delta\phi_\mathrm{GW}$, where $\phi_\mathrm{GW}$ is the ``carrier
phase'', defined by $\phi_{\rm GW} = 2 \phi$ with $\phi=\int\omega
\,dt$, and $\delta\phi_\mathrm{GW}$ is a standard precessional
correction, arising from the changing orientation of the orbital plane.
The precessional correction $\delta\phi_\mathrm{GW}$ can be computed by
standard methods using numerical integration, see Ref.~\cite{ACST94}.
Thus, the carrier phase $\phi_\mathrm{GW}$ constitutes the main
theoretical output to be provided for the templates, and can directly be
computed numerically from our main result, Eq.~\eqref{omdot}. In absence
of orbital-plane's precession, \textit{e.g.}, for spins aligned or
anti-aligned with the orbital angular momentum, the GW phase reduces to
$\phi_{\rm GW}$, and the latter can be obtained by integrating
\textit{analytically} Eq.~\eqref{omdot}. We get
\begin{eqnarray}\label{phiGW} 
\phi &&= \phi_0 -\frac{1}{32\nu}\,\left \{ x^{-5/2}+ x^{-3/2} \left (
\frac{3715}{1008}+ \frac{55}{12}\,\nu \right ) + \frac{x^{-1}}{G m^2}
\left ( \frac{235}{6} S^\text{c}_\ell + \frac{125}{8}\frac{\delta m}{m}
\Sigma^\text{c}_\ell\right ) \right . \nonumber \\ && - 10 \pi\,x^{-1} +
x^{-1/2}\left ( \frac{15293365}{1016064} + \frac{27145}{1008}\,\nu +
\frac{3085}{144}\,\nu^2 \right ) + \pi\,\ln x \left (
\frac{38645}{1344} - \frac{65}{16}\,\nu \right ) \nonumber \\ && \left
. + \frac{\ln x}{G m^2}\left [ \left ( -\frac{554345}{2016}-
\frac{55}{8}\,\nu \right ) S^\text{c}_\ell + \left ( -\frac{41745}{448}+
\frac{15}{8}\,\nu \right ) \,\frac{\delta m}{m} \Sigma^\text{c}_\ell \right
] \right \}\,,
\end{eqnarray}
where $\phi_0$ denotes some constant phase. In terms of the
single-spin variables $\mathbf{S}^\text{c}_1$ and
$\mathbf{S}^\text{c}_2$, the spin-dependent part of the above equation
reads
\begin{align}\label{phiGWs} 
\mathop{\phi}_\text{S} &= -\frac{1}{32\nu}\,\sum_{i = 1,2} \chi^\text{c}_i\,
\kappa^\text{c}_i \left \{ \left( \frac{565}{24}\,\frac{m_i^2}{m^2} +
\frac{125}{8}\nu \right) x^{-1} \right . \nonumber \\ & \qquad\left . +
\left [ \left ( -\frac{732985}{4032}- \frac{35}{4}\,\nu \right )
\,\frac{m_i^2}{m^2} + \left ( -\frac{41745}{448}+
\frac{15}{8}\,\nu \right ) \,\nu\,\right ] \ln x \right \}\,,
\end{align}
where $\chi^\text{c}_i$ and $\kappa^\text{c}_i$ are defined by
$\mathbf{S}^\text{c}_i=G
\,m_i^2\,\chi^\text{c}_i\,\hat{\mathbf{S}}^\text{c}_i$ and
$\kappa^\text{c}_i= \hat{\mathbf{S}}^\text{c}_i \cdot \bm{\ell}$. The
number of accumulated GW cycles between some minimal and maximal
frequencies is
\begin{equation}\label{NGW}
\mathcal{N}_\mathrm{GW} =
\frac{\phi_\mathrm{max}-\phi_\mathrm{min}}{\pi}\,.
\end{equation}
We list in Tables~\ref{Tabligo} and~\ref{Tablisa} the number of
accumulated GW cycles~\eqref{NGW} for typical binary masses in the
most sensitive frequency band of ground-based and space-based
detectors. For comparison we also show the contribution due to
spin-spin terms at 2PN order evaluated in Ref.~\cite{KWW93}, as well
as those due to the non-spin 3PN and 3.5PN orders computed
in~\cite{BFIJ02,BDEI04}. We denote $\xi^\text{c} =
\hat{\mathbf{S}}^\text{c}_1\cdot \hat{\mathbf{S}}^\text{c}_2$. From
the Tables~\ref{Tabligo} and~\ref{Tablisa} we deduce two important
results of this paper. First, we see that at 2.5PN order, if spins are
maximal, \textit{i.e.} $\chi^\text{c}_i=1$, the number of GW cycles
due to spin couplings is comparable to the number of GW cycles due to
non-spin terms.  Secondly, we find that for small mass-ratio binaries,
the number of GW cycles due to linear spins at 2.5PN order can be much
larger than the number of GW cycles due to spin-spin terms at 2PN
order. These results thus show that the 2.5PN spin terms evaluated in
the present paper have to be included in the GW templates if we want
to extract accurately the binary parameters.
\begin{table*}[h]
\caption{
  Post-Newtonian contributions to the number of GW
  cycles~\eqref{NGW} accumulated from $\omega_\mathrm{min} =
  \pi\times 10\,\mathrm{Hz}$ to $\omega_\mathrm{max} =
  \omega_\mathrm{ISCO}=1/(6^{3/2}\,m)$ for binaries detectable by
  LIGO and Virgo. For comparison, we add the contributions of
  spin-spin terms at 2PN order (we denote $\xi^\text{c} =
  \hat{\mathbf{S}}^\text{c}_1\cdot \hat{\mathbf{S}}^\text{c}_2$) and
  non-spin terms at 3PN and 3.5PN orders.
\label{Tabligo}}
\begin{center}
{\scriptsize
\begin{tabular}{|l|c|c|c|}\hline
& \multicolumn{1}{c|}{$(10+1.4)M_\odot$} &
\multicolumn{1}{c|}{$(10+10)M_\odot$} &
\multicolumn{1}{c|}{$(1.4+1.4)M_\odot$} \\ \hline\hline Newtonian &
$3577$ & $601$ & $16034$ \\ 1PN & $+213$ & $+59.3$ & $+441$\\ 1.5PN &
$-181 + 114\, \kappa^\text{c}_1\,\chi^\text{c}_1 + 11.8\,
\kappa^\text{c}_2\,\chi^\text{c}_2$ & $-51.4 + 16.0\,
\kappa^\text{c}_1\,\chi^\text{c}_1 + 16.0\,
\kappa^\text{c}_2\,\chi^\text{c}_2$ & $ -211 + 65.7\,
\kappa^\text{c}_1\,\chi^\text{c}_1 + 65.7\,
\kappa^\text{c}_2\,\chi^\text{c}_2$ \\ 2PN & $+ 9.8 - 4.4\,
\kappa^\text{c}_1\,\kappa^\text{c}_2\,\chi^\text{c}_1\,\chi^\text{c}_2
+ 1.5\, \xi^\text{c}\,\chi^\text{c}_1\,\chi^\text{c}_2$ & $+4.1 - 3.3\,
\kappa^\text{c}_1\,\kappa^\text{c}_2\,\chi^\text{c}_1\,\chi^\text{c}_2
+ 1.1\, \xi^\text{c}\,\chi^\text{c}_1\,\chi^\text{c}_2$ & $+ 9.9 - 8.0\,
\kappa^\text{c}_1\,\kappa^\text{c}_2\,\chi^\text{c}_1\,\chi^\text{c}_2
+ 2.8 \,\xi^\text{c}\,\chi^\text{c}_1\,\chi^\text{c}_2$ \\ 2.5PN & $-20 +
33.9\, \kappa^\text{c}_1\,\chi^\text{c}_1 + 2.9\,
\kappa^\text{c}_2\,\chi^\text{c}_2$ & $-7.1 + 5.7\,
\kappa^\text{c}_1\,\chi^\text{c}_1 + 5.7\,
\kappa^\text{c}_2\,\chi^\text{c}_2$ & $-11.7 + 9.3\,
\kappa^\text{c}_1\,\chi^\text{c}_1 + 9.3\,
\kappa^\text{c}_2\,\chi^\text{c}_2$ \\ 3PN & $+2.3$ & $+2.2$ & $+2.6$
\\ 3.5PN & $-1.8$ & $-0.8$ & $-0.9$ \\ \hline
\end{tabular}}\end{center}
\end{table*}
\begin{table*}[h]
\caption{Post-Newtonian contributions to the number of GW
  cycles~\eqref{NGW} accumulated until $\omega_\mathrm{max} =
  \omega_{\rm ISCO}=1/(6^{3/2}\,m)$ over one year of integration,
  for binaries detectable by LISA. For comparison, we add the
  contribution of spin-spin terms at 2PN order (we denote
  $\xi^\text{c} = \hat{\mathbf{S}}^\text{c}_1\cdot
  \hat{\mathbf{S}}^\text{c}_2$) and non-spin terms at 3PN and 3.5PN
  orders.
\label{Tablisa}}
\begin{center}
{\scriptsize
\begin{tabular}{|l|c|c|c|}\hline
& \multicolumn{1}{c|}{$(10^6+10^6)M_\odot$} &
\multicolumn{1}{c|}{$(10^6+10^5)M_\odot$} &
\multicolumn{1}{c|}{$(10^5+10^5)M_\odot$} \\ \hline\hline Newtonian &
$2267$ & $4985$ & $9570$ \\ 1PN & $+ 134$ & $+ 281$ & $+ 323$\\ 1.5PN
& $-92.4 + 28.8\,\kappa^\text{c}_1\,\chi^\text{c}_1 + 28.8\,
\kappa^\text{c}_2\,\chi^\text{c}_2$ & $-243 + 161\,
\kappa^\text{c}_1\,\chi^\text{c}_1 + 11.5\,
\kappa^\text{c}_2\,\chi^\text{c}_2$ & $-170 + 53\,
\kappa^\text{c}_1\,\chi^\text{c}_1 + 53\,
\kappa^\text{c}_2\,\chi^\text{c}_2$ \\ 2PN & $ +6.0 - 4.8\,
\kappa^\text{c}_1\,\kappa^\text{c}_2\,\chi^\text{c}_1\,\chi^\text{c}_2
+ 1.7\, \xi^\text{c}\,\chi^\text{c}_1\,\chi^\text{c}_2$ & $ + 12.5 - 4.4\,
\kappa^\text{c}_1\,\kappa^\text{c}_2\,\chi^\text{c}_1\,\chi^\text{c}_2
+ 1.5\, \xi^\text{c}\,\chi^\text{c}_1\,\chi^\text{c}_2$ & $ + 8.7 - 7.1\,
\kappa^\text{c}_1\,\kappa^\text{c}_2\,\chi^\text{c}_1\,\chi^\text{c}_2
+ 2.4\, \xi^\text{c}\,\chi^\text{c}_1\,\chi^\text{c}_2$ \\ 2.5PN & $ -9.0 +
7.1\, \kappa^\text{c}_1\,\chi^\text{c}_1 + 7.1\,
\kappa^\text{c}_2\,\chi^\text{c}_2$ & $-26.5 + 47.0\,
\kappa^\text{c}_1\,\chi^\text{c}_1 + 2.7\,
\kappa^\text{c}_2\,\chi^\text{c}_2$ & $ -11.0 + 8.7\,
\kappa^\text{c}_1\,\chi^\text{c}_1 + 8.7\,
\kappa^\text{c}_2\,\chi^\text{c}_2$ \\ 3PN & $+2.3$ & $+2.3$ & $+2.5$
\\ 3.5PN & $-0.9$ & $-2.3$ & $-0.9$ \\ \hline
\end{tabular}}\end{center}
\end{table*}

The number of accumulated GW cycles can be a useful diagnostic to
understand the importance of spin effects, but taken alone it provides
incomplete information. First, $\mathcal{N}_{\rm GW}$ is related only to
the number of orbital cycles of the binary within the orbital plane, but
it does not reflect the precession of the plane, which modulates the
wave form in both amplitude and phase. These modulations are important
effects. In fact, it has been shown~\cite{GKV03,BCV03b,GK03} that
neither the standard non-spinning-binary templates (which do not have
built-in modulations) nor the original Apostolatos templates~\cite{A95}
(which add only modulations to the phase) can reproduce satisfactorily
the detector response to the GWs emitted by precessing binaries.
Modulations both in the phase and the amplitude of the wave form must be
included~\cite{BCV03b,PBCV04,BCPV04,BCPTV05}. Second, even if two
signals have $\mathcal{N}_\mathrm{GW}$ that differ by $\sim 1$, one can
always shift their arrival times to obtain higher overlaps, but at the
cost of introducing systematic errors in the binary parameters. To
quantify the impact of the 2.5PN spin terms in detecting GWs from
spinning, precessing binaries, one should evaluate the maximized overlap
(fitting factor) between the 2.5PN template family and the 2PN template
family used in Refs.~\cite{BCV03b,PBCV04,BCPV04,BCPTV05}. Those template
families are defined by the GW signal computed along the binary
evolution together with the spin and angular momentum precession
equations. We expect that the maximized overlap between the 2.5PN and
2PN templates could be high, because the spins and the directional
parameters entering the template families provide much leeway to
compensate for non-trivial variations of the phasing (see \textit{e.g.},
Table~II in Ref.~\cite{PBCV04} where maximized overlaps between several
PN templates were computed). This study goes beyond the goal of this
paper and will be tackled in future work.

\section{Conclusions}\label{secIX}

Within the multipole-moment formalism developed in
Refs.~\cite{BD86,B87,BD89,B95,B96,B98mult}, we obtained the SO
couplings, 1PN order beyond the dominant effect, in the binary's mass
and current quadrupole moments, as well as in the GW energy flux. The
current-quadrupole moment with SO couplings at 1.5PN order was derived
in Ref.~\cite{OTO98}, but our result differs from the expression
computed there for two reasons. (i) The authors of Ref.~\cite{OTO98}
neglected the non-compact support terms which originate from the
non-linearities of the Einstein field equations and are not negligible
at this order. (ii) Their result for the compact-support terms is
affected by a computational error. The mass-quadrupole moment with SO
couplings at 2.5PN order (including all compact and non-compact support
terms) is computed here for the first time.

The binary's energy and the spin precession equations including SO couplings
through 2.5PN order were computed in paper~I. They were used to derive the
secular evolution of the binary's orbital phase through 2.5PN order in the
spins. We found that the 2.5PN terms give a relevant contribution to the
number of accumulated GW cycles within the binary's orbital plane. In
Tables~\ref{Tabligo} and~\ref{Tablisa}, we listed the number of GW cycles for
typical binaries detectable with ground-based and space-based detectors, such
as LIGO/Virgo and LISA. When spins are maximal, the SO contribution at 2.5PN
order is comparable to that of the non-spin part at the same 2.5PN order. For
some binary mass-configurations, the SO contribution at 2.5PN order can be
larger than that of SS couplings at 2PN order.

In order to extract accurately the parameters of maximally or mildly
spinning binaries with ground-based detectors of first generation,
having typical signal-to-noise ratio (SNR) of the order of 10, we expect
the spin corrections through 3.5PN order to be sufficient. With
space-based detectors having SNR of the order of $10^2 \mbox{--} 10^3$,
we would need \textit{a priori} to compute non-spin and spin corrections
at much higher PN order (for parameter estimation). For what concerns
the impact of the SO couplings at 2.5PN order on the actual detection,
it would be relevant to evaluate the maximized overlaps between
templates that include SO effects through 2.5PN order against templates
that include SO and SS effects through 2PN order. We anticipate that, at
the cost of introducing systematic errors in the estimation of the
binary parameters, the maximized overlaps could be high. In fact, the
binary and directional parameters may compensate variations in the PN
phasing.

For future applications, we listed in Sec.~\ref{secVII} the relevant
equations defining the spinning dynamics and the GW phasing in terms of
the constant spin variables. Such formulation is broadly used in the
literature to define spinning, precessing templates for compact
binaries~\cite{KWW93, K95, ACST94,A95,A96a,A96b,GKV03,BCV03b,PBCV04,
GK03,BCPV04,GIKB04,BCPTV05}.

Finally, we computed the contributions of the spin terms to the location of
the innermost circular orbit in the case of black-hole binaries. The results
for equal-mass objects with maximal spins are summarized in
Table~\ref{Tableico}. Spin couplings at 1.5PN and 2.5PN orders can give
significant contributions to the energy and frequency at the ICO (and nearby).

\acknowledgments 

We are grateful to Achamveedu Gopakumar and Gerhard Sch{\"a}fer for
pointing out to us that without calculating the secular effects to the
spin precession equations, i.e., $[\dot{\mathbf{S}}_1]_{\rm sec}$ and
$[\dot{\mathbf{S}}_2]_{\rm sec}$, it is not possible to derive from
the balance equation the expression of $\dot{\omega}/\omega^2$ and
compute the GW phasing $\phi (\omega,\mathbf{S}_1,\mathbf{S}_2,
\bm{\ell})$.  

\bibliography{BBF06spin_11-02-10.bib}

\end{document}